\documentclass[sn-mathphys,iicol]{sn-jnl}

\usepackage{soul}
\usepackage{caption}
\usepackage{subcaption}

\definecolor{mycolor}{RGB}{ 186, 219, 237}

\jyear{2023}%

\theoremstyle{thmstyleone}%
\theoremstyle{thmstyletwo}%
\theoremstyle{thmstylethree}%

\renewcommand{\thesection}{\Roman{section}} 

\renewcommand{\thesubsection}{\Alph{subsection}.}
\begin{document}

\title[Article Title]{Double-light-sheet, Consecutive-overlapping Particle Image Velocimetry for the Study of Boundary Layers past Opaque Objects}


\author[1]{\fnm{Shuangjiu} \sur{Fu}}

\author*[1]{\fnm{Shabnam} \sur{Raayai-Ardakani}}\email{sraayai@fas.harvard.edu}


\affil[1]{\orgdiv{Rowland Institute}, \orgname{Harvard University}, \orgaddress{\street{100 Edwin H. Land Blvd.}, \city{Cambridge}, \postcode{02142}, \state{MA}, \country{USA}}}




\abstract{Investigation of external flows past arbitrary objects requires access to the information in the boundary layer and the inviscid flow to paint a full picture of their characteristics. However, in laser diagnostic techniques such as particle image velocimetry (PIV), limitations like the size of the sample, field of view and magnification of the camera, and the size of the area of interest restrict access to some or part of this information. Here, we present the implementation of a variation on the two-dimensional, two-component (2D-2C) PIV to access flows past samples larger than the field of view of the camera. We introduce an optical setup to use one laser to create a double-light-sheet illumination to access both sides of a non-transparent sample and employ a Computer Numerically Controlled (CNC) carrier to move the camera in consecutive-overlapping steps to perform the measurements. As a case study, we demonstrate the capability of this approach in the study of the boundary layer over a finite-size slender plate. {We discuss how access to details of a macro-scale flow can be used to} explore the local behavior of the flow in terms of velocity profiles and the shear stress distribution. The boundary layers are not fully captured by the Blasius theory and are affected by a distribution of pressure gradient which in comparison results in regions of more attached or detached profiles. Ultimately, we show that the measurements can also be used to investigate the forces experienced by the body and decompose their effects into different components.

}




\maketitle
\section{Introduction}\label{sec1}

External flows past objects make up a substantial portion of flows that are of interest to fluid mechanics researchers and aero-/hydrodynamic applications. Exploration of these flows inherently requires access to both the viscous boundary layer (near-field) and the far-field inviscid flow. Idealized models of these flows assume the bodies of interest are suspended in a sea of fluid without any boundaries, have certain distributions of pressure gradients, or even very specific geometric boundaries. However, in real-life experimental scenarios in wind or water tunnels, flows are usually bound by near or far boundaries, the pressure distribution is not fully under the control of the operator \citep{liepmann1943investigations, Schlichting2014boundary}, and samples can come in complex geometric shapes without closed-form mathematical definitions \citep{vollsinger2005wind, pennycuick1996wingbeat}. This has thus resulted in discrepancies between reported measurements \citep{chauhan2009criteria}, especially in studies of boundary layers. Methods such as streamlining and sharpening of the leading and trailing edges \citep{grek1996experimental}, and careful surface adjustments or designs for control of pressure gradients \mbox{\citep{liepmann1943investigations,  bross2019interaction}} have proven to be challenging but promising in recreating some of these cases in laboratory scales. 

In recent decades, laser diagnostic techniques such as particle image velocimetry (PIV) \mbox{\citep{adrian2011particle}} have greatly advanced our experimental toolboxes to gain better measurements and understanding of flow fields. {However, PIV measurements can also be limiting in the extent of the information that can be gathered within one experiment. For example, the field of view of the camera and magnification used in the imaging defines the extent of the region of investigation \mbox{\citep{michalek2022influence}}, and at times can only be limited to a high-resolution view of the boundary layer \mbox{\citep{abu_rowin_ghaemi_2019}}, or a lower resolution view of a wide area in the flow incorporating more details of the far-field information \mbox{\citep{terra2016drag}}. In addition, refractive index-matching is not always a viable option, and non-transparent objects can place portions of the flow in shadows \mbox{\citep{kim2015experimental, nair_kazemi_curet_verma_2023, du2022control}}. This can restrict studies of asymmetric flows or samples. Even for symmetric samples, it adds additional uncertainties to the measurements if the assumption of symmetry is not fully met in the experimental setup. }

Here, we implement a cost-effective variation of the 2D-2C PIV procedure with simultaneous 2-axis load measurements to study the flow field in both the boundary layer and the inviscid far field of an arbitrary opaque sample at high resolutions. To achieve this, we use a single laser and double light-sheet illumination strategy, combined with a consecutive-overlapping image acquisition supported by a single camera maneuvered by a Computer Numerically Controlled (CNC) stage. {We will discuss why multi-camera setups with overlapping fields of view (which are used for large fields of view \mbox{\citep{parikh2023lego, carmer2008evaluation, knopp2015investigation}}) would not be a feasible replacement for this approach} with the currently available high-speed cameras and lenses and might become possible if technological advances in digital high-speed photography and optics would allow the size of the cameras and lenses to shrink substantially. 

{To demonstrate the capability of this setup, we focus on the case of a laminar boundary layer past a finite-length flat plate. With the new advances and developments in the areas of unmanned aerial/underwater vehicles \mbox{\cite{floreano2015science, di2020bioinspired}} which are smaller than conventional vehicles and operate at lower speeds, there is an increased need for better ways of characterizing boundary layers over smaller bodies in moderate Reynolds numbers. In the range of Reynolds numbers studied here, the flow does not transition to turbulent but the wake past the trailing edge turns turbulent, and due to the limited length, the assumption of the asymptotic behavior of the boundary layer theory does not fully hold. }

{In addition, boundary layers over flat plates serve as references for comparison in studies of textures \mbox{\citep{grek1996experimental, walsh1984optimization, Bechert2000, raayai2017drag, raayai2019geometric, vukoslavcevic1992viscous, du2022control}}, roughness elements \mbox{\citep{kim2015experimental, michalek2022influence}}, or super-hydrophobic surfaces \mbox{\citep{xu2021superhydrophobic}}. To date, there are only a few studies that present experimental measurements of zero pressure gradient laminar boundary layers in the asymptotic region of a flat wall \mbox{\cite{Schlichting2014boundary, liepmann1943investigations, chauhan2009criteria}} and there are no experimental studies presenting the whole view of the flow around the entirety of a streamlined plate, with enough resolution to have the ability to analyze the details of the flow both in the far-field and the boundary layer without depending on any assumptions regarding the sample orientation or symmetry. Among the experimental studies available on boundary layers, strategies such as focusing on partial locations within the length of the plate \mbox{\citep{grek1996experimental, xu2021superhydrophobic, bross2019interaction}}, single side measurements \mbox{\citep{grek1996experimental, knopp2015investigation}}, or installation of a sample as part of the wind/water tunnel's wall \mbox{\citep{walsh1984optimization, Bechert2000, abu_rowin_ghaemi_2019}} have been previously considered.}

This paper is organized in the following manner: in Sec. \ref{ExpSetup} we discuss the experimental facility and the implemented procedure. In Sec. \ref{Results} we present the results of the experiments performed on a slender, finite-length flat plate sample with a streamlined leading edge; In Sec. \ref{Results} \ref{contours} and Sec. \ref{Results} \ref{angle_attack} we cover the details of the far-field of the flow, and in Sec. \ref{Results} \ref{BL} compare and contrast the data against the first order boundary layer theory as described by Prandtl and Blasius \citep{Schlichting2014boundary}, and ultimately show in Sec. \ref{Results} \ref{shear} and Sec. \ref{Results} \ref{all_force} that the flow, local shear stress distribution, and load measurements can be used to decompose the total forces experienced by the sample into various components.

\begin{figure*}
\centering
\includegraphics[width = \textwidth]{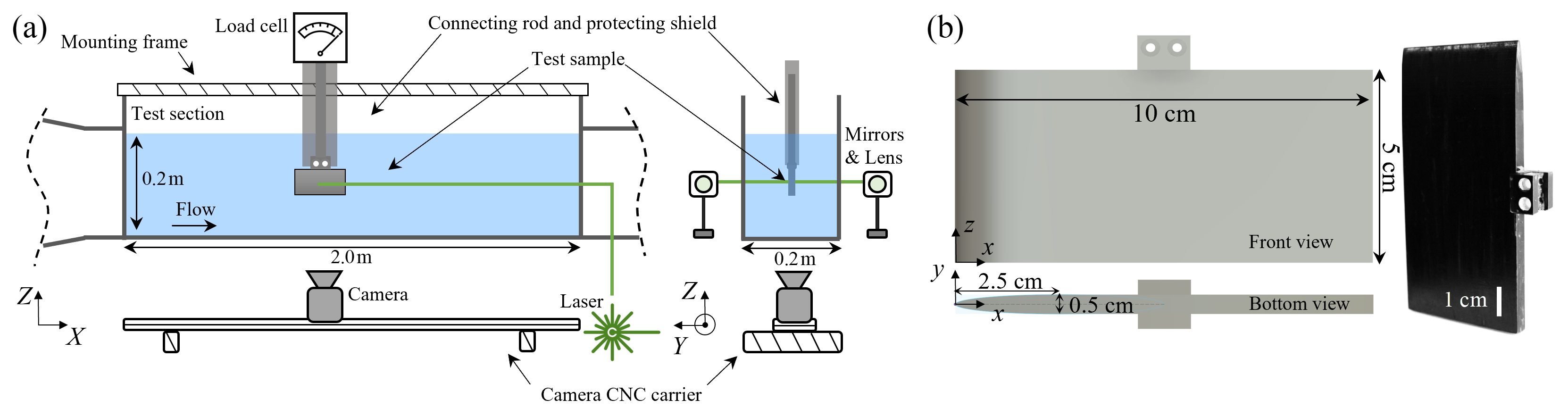}
\caption{(a) Schematic of the experimental water tunnel facility and the PIV components (front view and side view), with an active cross-sectional area of $0.2 \times 0.2 \ {\rm m}^2$ and $2 \ {\rm m}$ length. The global coordinate system shown as $(X, Y, Z)$ is used during the experimental procedure to control the location of the sample and camera. (b) Schematic of the slender sample (front and bottom views) with elliptic leading edge and the top handle used for connection to the load cell (left) and an image of the actual sample (right). Local $(x,y,z)$ directions are used in the analysis of the results. {The scale bar on the bottom right side of the sample is 1 cm}.}

\label{fig:detail.jpg}
\end{figure*}

\section{Experimental method}\label{ExpSetup}
\subsection{Flow facility and sample of interest} \label{facility}

The experiments are performed in a $2 \ {\rm m}$ long water tunnel with a rectangular cross-section of $20 \times 27 \ {\rm cm}^2$ where the water height is kept at $20 \ {\rm cm}$ during the experiments (see Fig. \ref{fig:detail.jpg}(a)). Experiments are performed at three free-stream velocities less than $0.25 \ \rm{m/s}$ ($0.122$, $0.185$, and $0.242 \ {\rm m/s}$) where the turbulence intensity of the free stream is less than or about $1\%$. The free-stream velocity is controlled and set via the main computer and an analog output through a Data Acquisition (DAQ) system (NI DAQ USB-6001) connected to the tunnel. A separate flow meter measures the flow rate of the pump and is set to communicate with the main computer via an analog input of the DAQ.

The sample of interest is a slender, short, symmetric plate of $100 \  {\rm mm}$ long ($L$), $50 \  {\rm mm}$ wide ($b$), and $5 \  {\rm mm}$ in thickness ($h$) with a streamlined elliptical leading edge (see Fig. \ref{fig:detail.jpg}(b)) and is fabricated using 3D printing (Formlabs Form3B 3D printer and colored photo-polymer resin). The leading edge of the sample within $0\leqslant x \leqslant 25 \ {\rm mm}$ is streamlined in an elliptic form with a $1/10$ ratio of the semi-minor and semi-major axes and past that the two sides of the sample in $25\leqslant x \leqslant 100 \ {\rm mm}$ are flat, ending at a blunt trailing edge. The PIV measurements are performed in the middle of the sample ($z \approx 2.5 \ {\rm cm}$) to reduce the effect of the top and bottom boundaries of the sample.

Using a connecting rod on the top, the sample is connected to a 2-axis load cell consisting of Linear Variable Differential Transformers (LVDT), and suspended in the tunnel at a distance of $76\ {\rm cm}$ from the tunnel entrance and slightly lower than the mid-height of the tunnel in the $Z$ direction. As seen in Fig. \mbox{\ref{fig:detail.jpg}}(a), {to avoid any effects of gravity waves \mbox{\cite{kundu2015fluid}} that could exist at the free-surface}, the length and thickness of the sample are in the horizontal plane and the width of the sample is in the $Z$ direction. The motor and the pump are separated from the experimental area via damping rubber cushions isolating the two from the main experimental area. The rod connecting the sample to the load cell is protected by a streamlined shield to minimize the impact of the rod on the load measurements and avoid unwanted wakes behind the connecting rod and above the sample. The load cell is set to communicate with the main computer via two of the analog inputs of the DAQ system.

\begin{figure*}
\centering
\includegraphics[width = \textwidth]{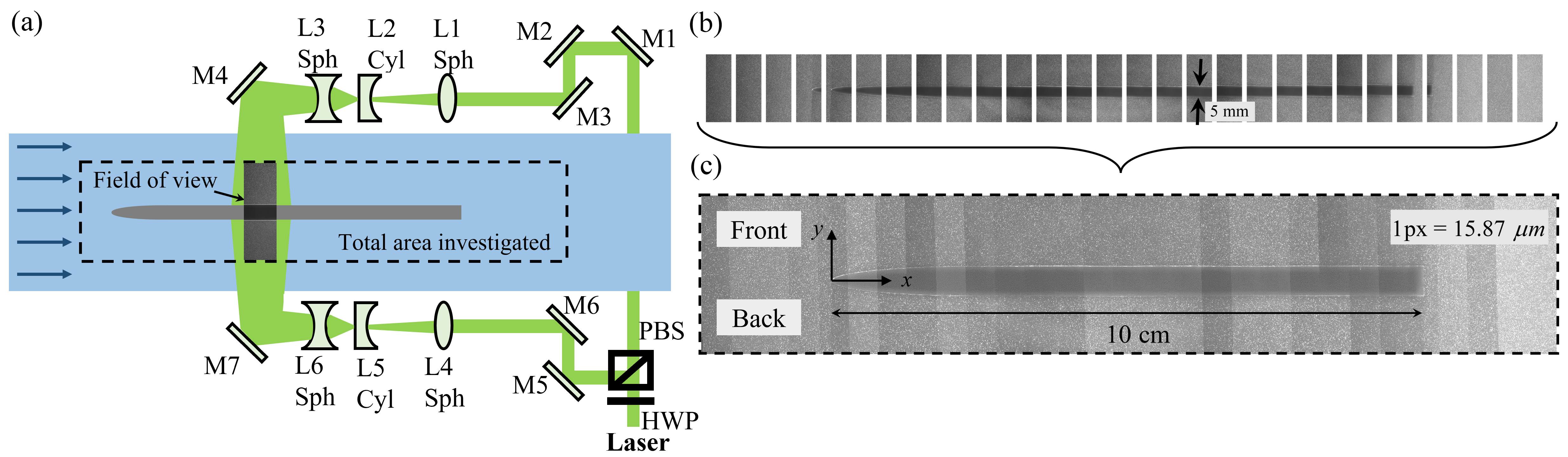}
\caption{(a) Schematic of a double-sheet setup for simultaneous illumination of two sides of an opaque sample. The laser beam is divided into two beams through an HWP and a PBS and is guided through mirrors toward two sets of identical light-sheet optics. (b) Snapshots of a series of images acquired using the consecutive-overlapping technique covering the entire length of the sample as well as before and after the sample. (c) The fully stitched view of the images from part (b) showing the full view of the sample. The region of flow in $y>0$ is denoted as ``Front'' and the region in $y<0$ is denoted as ``Back'' throughout the text. The starting point of the leading edge is the local origin of the $(x,y)$ coordinate used.}
\label{fig:Schematic.png}
\end{figure*}

\subsection{2D-2C PIV}
The velocity field is measured using a 2D-2C PIV procedure. The setup consists of a double-pulsed Nd:YAG laser (Evergreen EVG00200, Quantel Laser) operated at $15 \ {\rm Hz}$ repetition rate and nominal output energies of $10 \ {\rm mJ}$ or $20 \ {\rm mJ}$ per pulse for different free stream velocities, a high-speed camera (Chronos 2.1, Kron Technologies Inc.) at a resolution of 720 $\times$ 1920 pixels with a 100 mm macro lens (Canon EF 100mm f/2.8L Macro Lens), and a timing unit (Arduino Teensy Board) with an Arduino program similar to Ref. \cite{teensy} uploaded on the board and used for synchronizing the instances of laser pulses and camera capture. The timing between the two laser/camera shots is set using an analog output from the main computer to the timing unit via a DAQ output. For velocities less than $0.2 \ {\rm m/s}$ the timing is set at $\delta t = 1000 \ \mu {\rm s}$ and for free-stream velocity of $0.242 \ {\rm m/s}$ the timing is set to $\delta t = 900 \ \mu {\rm s}$. The camera is situated underneath the tunnel and its location is automatically controlled using a CNC motorized stage in all three directions. Water is seeded with $10 \ \mu {\rm m} $ hollow glass particles (TSI incorporated).

To access the velocity field on both sides of the opaque sample with only one light source, we use a double light-sheet strategy (as opposed to using multiple light sources \citep{gehlert2023vortex}) as illustrated in Figs. \ref{fig:detail.jpg}(a) and \ref{fig:Schematic.png}(a). In this method, various optical elements are configured so that the incoming linearly polarized laser beam is divided into two beams using a half-wave plate (HWP) and a polarizing beam-splitter (PBS) and directed toward the Front and Back of the tunnel via multiple mirrors (M1-3, M5-6, Thorlabs Nd:YAG Mirrors, 524 - 532 nm) where two lens combinations (2 spherical (L Sph) and one cylindrical (L Cyl)) are used to create light sheets (about $1\ {\rm mm}$ thick) to illuminate the Front and Back sides of the sample. L1 and L4 are spherical lenses with $+300$ mm focal lengths, L2 and L5 are cylindrical lenses with $-50$ mm focal lengths, and L3 and L6 are spherical lenses with $-100$ mm focal lengths. These two light sheets are parallel to each other and in the absence of the sample, the two would meet to create a sheet with nearly double the illumination intensity. The field of view of the camera thus captures the flow on both sides of the sample without any shadows (Fig. \ref{fig:Schematic.png}). The use of beam-splitters and single light source have previously been employed for 3D, multi-plane, and holographic PIV measurements as well \mbox{\citep{arroyo2008recent, sheng2003single, ganapathisubramani2005dual}}.

To access the boundary layer (near-wall) information, the imaging magnification is set to each pixel capturing 15-16 $\mu {\rm m}$ ($1 \ {\rm mm} \cong$ 63-64 $ {\rm px}$). As a result, the field of view of the camera is limited to about 11.5 mm of the sample (720 px, in the streamwise direction) at a time while the length of the total area of interest (sample,  before the leading edge, and after the trailing edge) is about 180 mm. Thus, to image the whole sample, using the CNC stage, the camera is swept in consecutive-overlapping steps (about $40-50\%$ overlap) covering the entire length of the sample as well as a few steps before the leading edge and after the trailing edge (Fig. \ref{fig:Schematic.png}). {Note that the overlap here is chosen conservatively and lower overlaps are all valid choices}. {This motorized system allows us to control the displacement of the camera with $0.1 \ {\rm mm}$ accuracy in all three directions (as noted by the manufacturer's specification of the CNC rails and motors).} Larger robotic/motorized PIV systems are also being considered for large field-of-view measurements and moving objects in industrial applications \mbox{\citep{michaux2018robopiv}}. The light sheet optics (Lenses L1-6 in Fig.\ref{fig:Schematic.png}(a)) are manually moved to illuminate the respective fields of view. Each field of view captures about 25 ${\rm mm}$ in the $y$ direction (about 12.5 ${\rm mm}$ on either side of the center-line of the sample ($y=0$) which is sufficient to extract the boundary layer information and have access to the inviscid far-field information. {For larger areas of interest and samples, the camera needs to be swept in both directions \mbox{\cite{fu2023multisheet}.}} While only 28 overlapping frames are shown in Fig. \ref{fig:Schematic.png}(b), the experiments presented here are repeated over 36 overlapping steps.

\begin{figure}
    \centering
    \includegraphics[width = 0.45\textwidth]{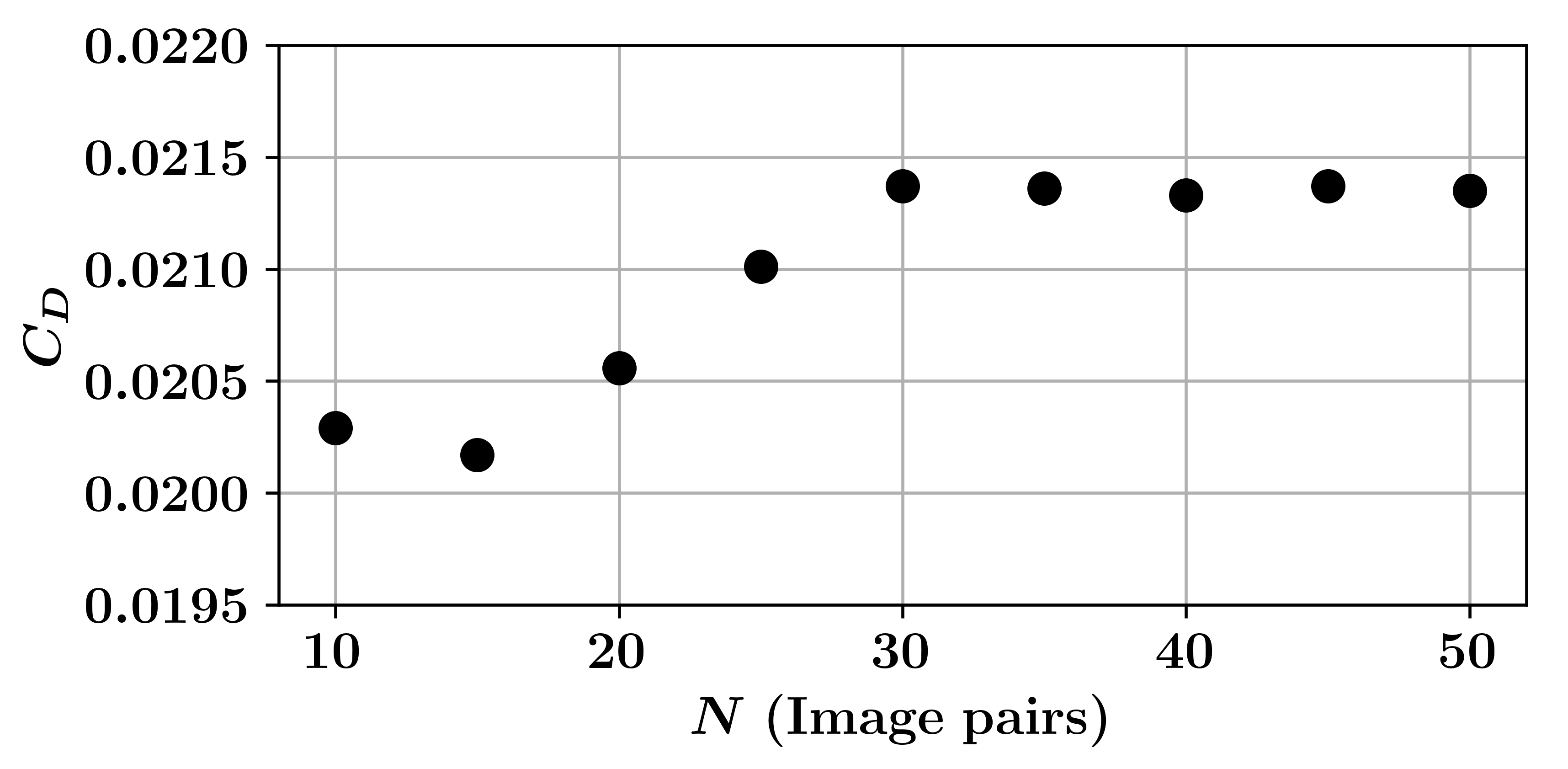}
    \caption{Drag coefficient calculated from a control volume analysis (more details in Sec. \ref{Results} \ref{all_force}) for a sample at ${\rm Re}_L = 12,200$ as a function of the number of image pairs used in the PIV analysis.}
    \label{fig:CD_N}
\end{figure}

At each location, 50 image pairs are captured and grouped together as a function of the global location of the left edge of the images. It can be shown that 30 image pairs per single capture are enough for the mean and variance of the measurements to converge, and for extra caution, here 50 image pairs have been used. In addition, due to close to $50\%$ overlaps between two consecutive steps, the total available image pairs capturing each local scene is 100 with 50 pairs from two consecutive experiments. This converging behavior transfers to all the other calculated variables as well and for example Fig. \mbox{\ref{fig:CD_N}} shows the convergence of the drag coefficient calculated using a control volume (details of the analysis presented in Sec. \mbox{\ref{Results} \ref{all_force}}) as a function of the number of image pairs used {for a sample at ${\rm Re}_L = \rho U_{\infty}L/\mu = 12,200$}. Afterwards, the global locations and the physical size of the pixels are used to stitch the images together to form the view of the entire sample. An example of the series of overlapping images and the final stitched view are shown in Figs. \ref{fig:Schematic.png}(b) and \ref{fig:Schematic.png}(c). Throughout the text, regions of flow in $y>0$ are denoted as ``Front'' and regions in the $y<0$ area are noted as ``Back''. 

\begin{figure*}
    \centering
    \includegraphics[width = 1\textwidth]{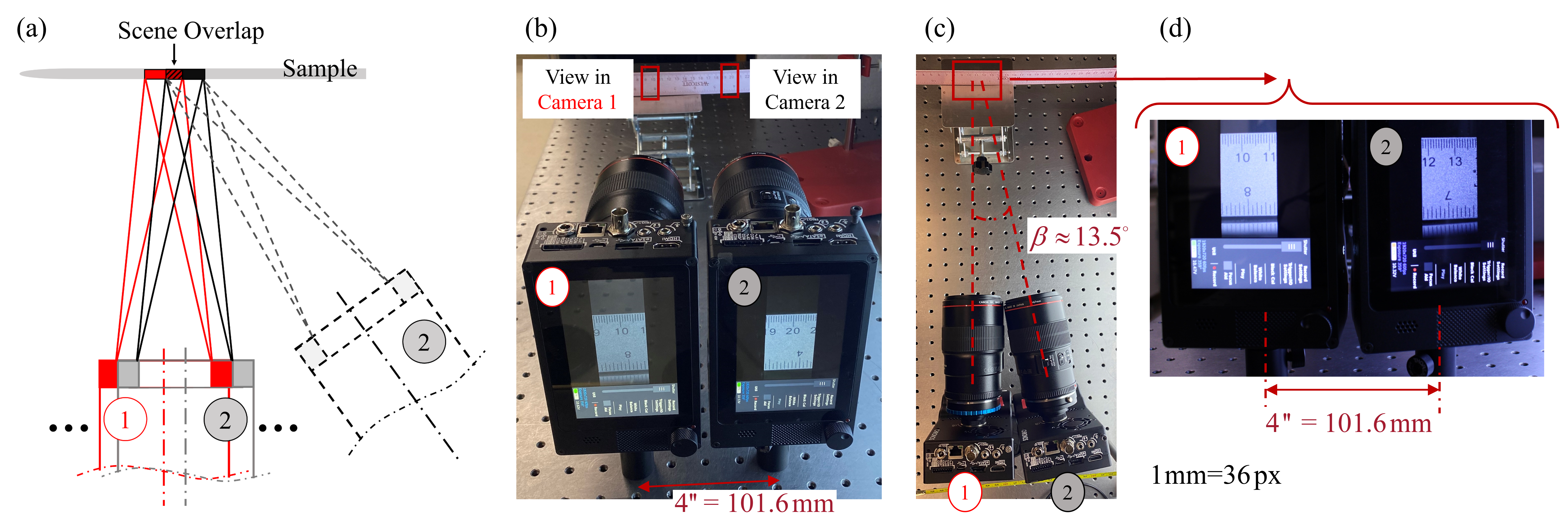}
    \caption{(a) Schematic of the location of two cameras (to scale) capturing the view of two overlapping scenes in consecutive-overlapping imaging. The red camera 1 captures the red view and the grey camera 2 captures the black view. The lens opening is slightly smaller than the outer radius of its body. The dashed grey camera represents a hypothetical titled camera viewing the black scene. (b) Reconstruction of the case of two side-by-side cameras with parallel focal axes viewing a ruler (object). The screens show the same separation as the center-to-center distance of the two cameras. (c) Reconstruction of the same two cameras with the second camera rotated to the maximum possible to capture a scene closer to the view of camera 1. (d) View of the camera screens from the setting of part (c).}
    \label{fig:side_by_side}
\end{figure*}

\subsection{Consecutive-overlapping imaging vs. simultaneous multi-camera alternative}

While in theory, consecutive-overlapping imaging can be recreated by a simultaneous multi-camera imaging scheme, the physical sizes of the available high-speed camera and macro lens limit the possibility of its execution at the resolution used here. As far as we know, previous investigations employing multiple cameras simultaneously for 2D-2C PIV have used lower spatial resolutions of for example 1 ${\rm mm} \cong$ 3.5 ${\rm px}$ \mbox{\citep{carmer2008evaluation}}  or 1 ${\rm mm} \cong$ 6 ${\rm px}$ \mbox{\citep{knopp2015investigation}} and at a relatively larger distance from the imaging planes. At the resolution of 1 ${\rm mm} \cong$ 63-64 $ {\rm px}$ used here, capturing two overlapping scenes simultaneously requires two cameras to be placed side-by-side with their parallel focal axes at a distance of $5 \ {\rm mm}$ (Fig. \ref{fig:side_by_side}(a)). However, the dimensions of the camera in this work (96 ${\rm mm} \times$155 ${\rm mm} \times$67.3 ${\rm mm}$ in $X$, $Y$, and $Z$ directions) restrict the placement of a second camera in the $X$ direction to at least $96 \ {\rm mm}$ to the side of the first one. A view of two side-by-side, parallel cameras with a gap of about 6 $\rm mm$, fixed to an Imperial optical table, is shown in Fig. \mbox{\ref{fig:side_by_side}}(b) (in the air, with no air/glass/water interface) where, as expected, the separation between the two views of the object (ruler) is close to 101.6 ${\rm mm}$ as visible in the screens.

To close the spatial gap between the fields of view visible in the two cameras, we can place the second camera at an angle with respect to the first one (camera marked by dashed lines in Fig. \ref{fig:side_by_side}(a)). However, the dimensions of the camera and lens ($77.7 \ {\rm mm}$ in radius and $123 \ {\rm mm}$ in length) restrict the available physical space for the two to rotate. For example, using the same setup and fixed separation as the one in Fig. \mbox{\ref{fig:side_by_side}}(b) and rotating camera 2 to the maximum extent possible, as seen in Figs. \mbox{\ref{fig:side_by_side}}(c) and \mbox{\ref{fig:side_by_side}}(d), the view of camera 1 (focal axis perpendicular to the imaging plane) captures the region between $9.2\ {\rm cm}$ and $11.2 \ {\rm cm}$ of the ruler (distance of $\sim 20 \ {\rm mm}$) while camera 2 at an angle of $\beta \approx 13.5^{\circ}$ captures the span of $11.7 \ {\rm cm}$ and $13.8 \ {\rm cm}$ (distance of $\sim 21 \ {\rm mm}$), missing $5 \ {\rm mm}$ of the object and at a slightly closer distance (1 ${\rm mm} \cong$ 51-52 $ {\rm px}$) the two cameras miss almost $2.5 \ {\rm cm}$ of the view. Thus, to capture an overlapping scene both the separation of the two cameras and the angle between them needs to increase.

If more cameras were to be added, the new cameras would require even larger angles to capture a field of view close to the one seen by their neighbors and additional geometric restrictions arise. The added tilt angles result in an increase in the distance between the imaging and the lens planes which requires adjusting the focus and reducing the magnification. In addition, the tilted cameras will need Scheimpflug adapters to correct for any defocusing and added steps of calibrations and corrections to map the image pixels to the physical space.

Lastly, due to the air/glass/water interface in our experimental setup, care has been taken to make sure that the CNC rail, the camera, and the side edges of the sample are parallel to each other and to the bottom wall of the tunnel and by aligning the laser sheet horizontally, we are able to capture the flow field in a cross-sectional plane of the sample parallel to the glass wall and the camera without suffering from air/glass/water refraction. However, in a multi-camera alternative, the cameras that are placed at an angle need to have an accompanying prism at the air/glass interface to correct the effect of the light refraction, and with the intended overlap, the physical geometry of one or many of the prisms will restrict the space available for the others and can block the view of the neighbors.

Thus, besides the financial advantage, with all the above factors and the available hardware, the use of 36 simultaneous cameras is not a feasible replacement for recreating consecutive-overlapping imaging at the resolution of interest in this work. {In case of the availability of additional cameras, a combined multi-camera consecutive-overlapping procedure with two parallel cameras with non-overlapping fields of view (such as in Fig. \mbox{\ref{fig:side_by_side}(b))} is a feasible way to expedite the experimental procedure.}

\begin{figure*}[!ht]
    \centering
    \includegraphics[width = 1 \textwidth]{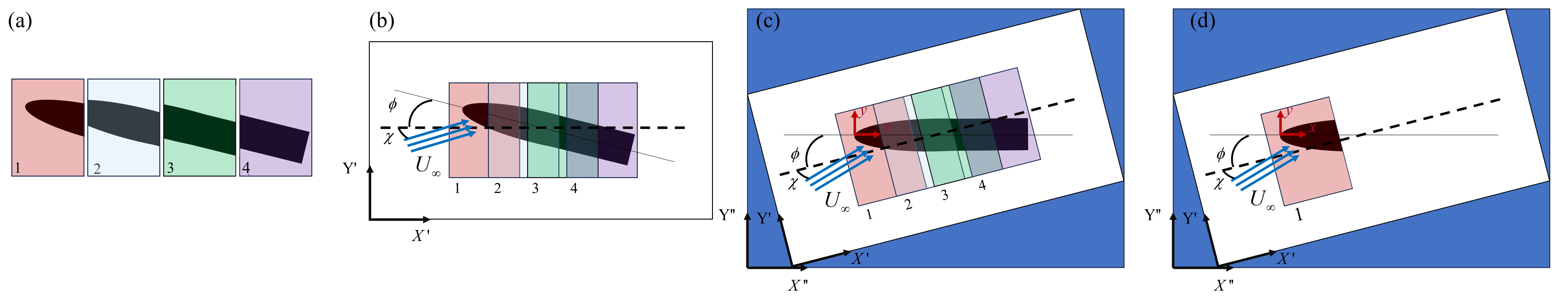}
    \caption{{Schematic of a hypothetical example of the stitching process used for the PIV analysis: (a) Separate images of a sample taken with about 40\% overlap, placed side by side. (b) The images are placed on a larger empty canvas according to their global locations known based on the overlaps and the physical scale of the pixels and identifying the center-line of the sample. The coordinate of the camera, and centerline of the sample, and the freestream direction are not necessarily parallel to each other. (c) The entire canvas is rotated so that the centerline of the sample is horizontal. This creates a larger canvas (blue area) so that the new image is saved in a 2D array on the computer. (d) Groups of images with their new locations are processed using the PIV algorithm and the coordinates of the PIV windows are translated from the global $(X'', Y'')$ to the $(x,y)$ coordinate system. Note that $\phi$ is the angle between the centerline and the coordinates of the images and $\chi$ is the angle between the freestream and the coordinates of the images.} }
    \label{fig:stitch_steps}
\end{figure*}

\subsection{Experimental procedures}

The main computer controlling the experimental procedure is set to communicate with the various components either through the DAQ (as discussed earlier) or directly via USB interfaces. After the sample and load cell are installed and secured, the free-stream velocity of the tunnel is set and the flow is left to reach a steady state. {Then the two light sheets are aligned on both sides of the tunnel and since the depth of field of the macro Lens is less than 1 mm, any potential misalignment between the two light sheets is visible in the camera and is adjusted when observed. To reduce the reflections from the sample, the sample is painted with black ink used in Sharpie\textsuperscript{\textregistered} markers prior to installation.} Then we set the timing $\delta t$ to synchronize the laser and camera according to the chosen free-stream velocity and the magnification. As per the camera manual, the frame rate of the camera is set to be slightly larger than $1/\delta t$ for the hardware to be able to process the captured signals from the timing unit properly. The CNC camera stage is driven via a USB interface and the open source package Open Builds CONTROL \citep{openbuilds} to move the camera to the location of interest. At this point, everything is ready for each set of experiments. Then using the multi-processing capabilities of the computer, the load measurement and camera capture are set to take place simultaneously. The entire measurement procedure is controlled via an in-house Python script, sending the signal to start and stop the experiments. After the experiments, the same Python script directs the acquired images and data to be saved in their appropriate locations. This procedure is then repeated for each imaging location in consecutive-overlapping steps as shown in Fig. \ref{fig:Schematic.png}(b).\

\subsection{Post-processing}

{To stitch the groups of images for post-processing, we create an empty canvas (2D array) and place all the images according to their global locations, overlaps, and the physical scale of the pixels. A schematic example is shown in \mbox{Fig. \ref{fig:stitch_steps}(a) and Fig. \ref{fig:stitch_steps}(b)}. Then using the Canny edge detection algorithm \mbox{\cite{canny1986computational}} we identify the boundaries and the centerline of the sample and we calculate the angle between the centerline and the horizontal direction of the stitched image ($X'$). Then, using this stitched image (Fig. \mbox{\ref{fig:Schematic.png}}(b)), the view of the sample is rotated (schematic in \mbox{Fig. \ref{fig:stitch_steps}(c)}) and fitted with the mathematically defined contour (here an elliptic leading edge, and the flat boundaries). 

Afterward, knowing the location of each of the batches of the images on this larger canvas, we go back to each group of the images and perform the PIV analysis only on the region of the rotated image as seen in \mbox{Fig. \ref{fig:stitch_steps} (d)}, and identify the locations of the centers of the PIV windows with respect to the origin of the rotated larger canvas $(X'', Y'')$. Note that the angles in the schematics of \mbox{Fig. \ref{fig:stitch_steps}} shown here are overly exaggerated and in the experiments shown here the angle $\phi$ is found to be 0.1$^{\circ}$.  Then the starting point of the leading edge is set as the origin of the local $x$ direction, and the zero in the local $y$ direction is set on the plane of symmetry of the sample. The mathematical relationship between the $(X '', Y'')$ and $(x,y)$ is used to translate the locations of the PIV windows to the coordinate of choice $(x,y)$. As a result, the flat boundaries of the sample are located at $y = \pm 2.5 \ {\rm mm}$ and parallel to the $x$ axis. (Not necessarily parallel to the free-stream velocity as discussed later in Sec. \mbox{\ref{Results} \ref{angle_attack}} and as shown in the schematic of \mbox{Fig. \ref{fig:stitch_steps}.}) This way, the exact wall-normal ($\hat{n}$) and tangential directions ($\hat{t}$) can be identified to allow for further analysis of the boundary layer information in \mbox{Sec. \ref{Results} \ref{BL}} }

\begin{figure*}[!ht]
    \centering
    \includegraphics[width = 0.98 \textwidth]{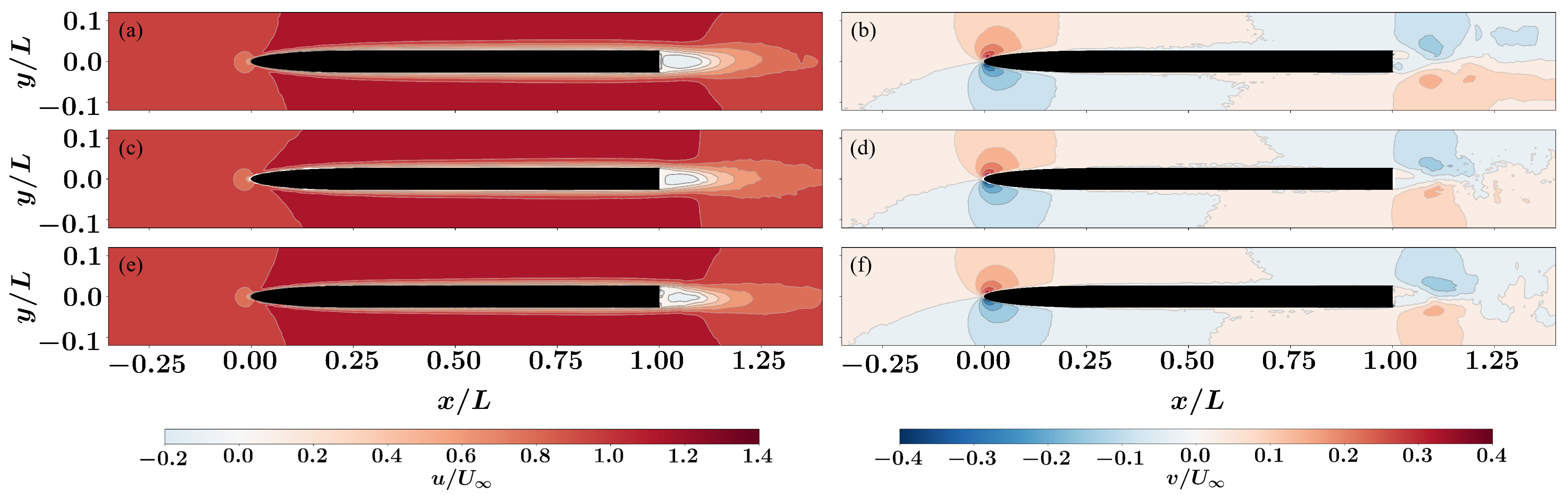}
    \caption{Contours of mean velocities $u$ (left) and $v$ (right) normalized by the free-stream velocity $U_{\infty}$ at global Reynolds numbers (a,b) ${\rm Re}_L = 12,200$, (c,d) ${\rm Re}_L = 18,500$, and (e,f) ${\rm Re}_L = 24,200$.}
    \label{fig:abcd}
\end{figure*}

The PIV images are processed with an in-house Python script partly using the open-source software OpenPIV \citep{alex_liberzon_2020_3911373} and an additional in-house correction loop for close-to-the-wall regions. The portion of the image pairs slightly far from the wall (further than 64 pixels away) are analyzed using functions from the OpenPIV package with $32 \times 32$ windows and a search area of $64 \times 64$ with $85\%$ overlap ({separation of about 63 $\mu \rm m$ between vectors}).  However, due to the large shear rate close to the wall boundary, to avoid bias errors close to the walls \citep{kahler2012uncertainty}, the first 64 pixels in the near-wall region are analyzed with an in-house cross-correlation scheme with a rectangular window of $32\times 16$ (smaller height in the normal direction) to reduce the averaging effects of square windows. {In all experiments, the timing $\delta t$ is chosen in a way that the fastest particles displace at a maximum of half of the window size minus one pixel to ensure that the slowest particles close to the wall have enough time to displace at least one pixel.}

After calculating the velocities for each experimental step, the mean velocity of each step and the velocity fluctuations of each instance are calculated and used to find the turbulent kinetic energy and Reynolds shear stress of each step. The instantaneous velocities are used to calculate the instantaneous vorticity and later the mean vorticity for each step is calculated from these instantaneous vorticities. {Ultimately all the results are stitched together based on the mapping found from the stitched images. In all the overlapping areas, an average of the results from both of the consecutive steps are used. Besides the boundary layer and shear stress analysis discussed in Sec. \mbox{\ref{Results} \ref{BL}} and \mbox{\ref{Results} \ref{shear}} which are performed for each step separately, in the rest of the paper the presented variables are averages of the means of two consecutive-overlapping steps.}

\section{Results and discussion} \label{Results}
\subsection{Velocity fields} \label{contours}

\begin{figure}[!ht]
    \centering
    \includegraphics[width = 0.48\textwidth] {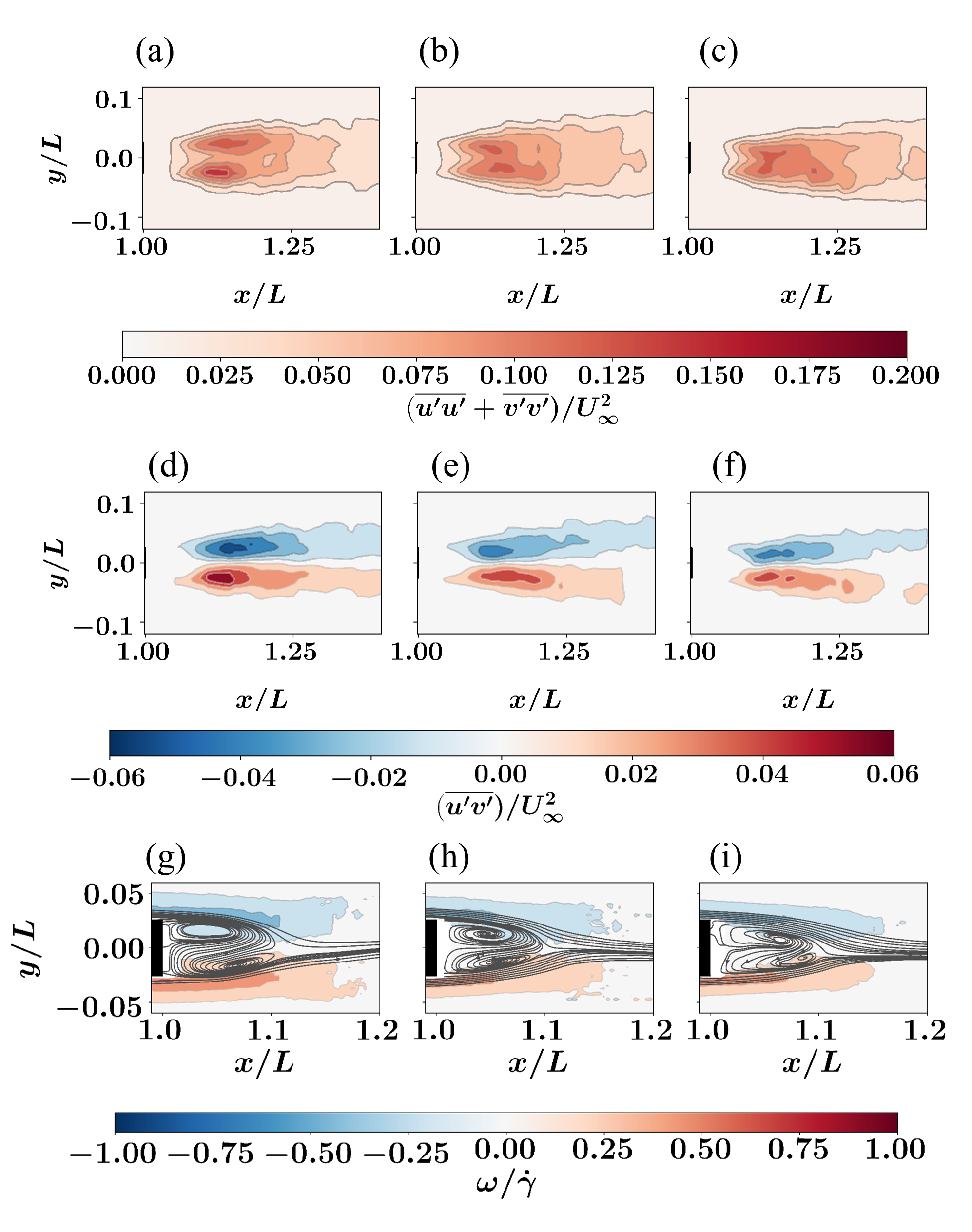}
    \caption{Contours of normalized turbulent kinetic energy (top row), Reynolds shear stress (middle row), and vorticity normalized by reference $\dot{\gamma}$ (bottom row) for Reynolds numbers of (a,d,g) ${\rm Re}_L = 12,200$, (b,e,h) ${\rm Re}_L = 18,500$, and (c,f,i) ${\rm Re}_L = 24,200$ past the trailing edge of the sample. Streamlines of the flow in the vicinity of the wall and the wake are also overlaid on top of the vorticity contours to show the extent of the separation bubble behind the sample.}
    \label{fig:fluctuations}
\end{figure}

The results of the mean velocity $u$, and $v$ (velocities in $x$ and $y$ directions respectively) around the entire sample, including both near- and far-field are presented in contour plots shown in Fig. \ref{fig:abcd} (normalized by the free-stream velocity) for three Reynolds numbers ${\rm Re}_L = 12,200$, ${\rm Re}_L = 18,500$, and ${\rm Re}_L = 24,200$. Here the global Reynolds number, ${\rm Re}_L = \rho U_{\infty}L/\mu$ is defined based on the total length of the sample and the free-stream velocity $U_{\infty}$, and $\rho$ and $\mu$ are the density and viscosity of water respectively. In all cases the region of the stagnation point at the leading edge is visible and due to the finite thickness of the sample, the flow in the inviscid area of either side is faster than the free-stream velocity. Overall the three cases have similar contour distributions in both directions and no clear difference is visible among the normalized velocities of various cases. 

As is expected from the ranges of the Reynolds numbers tested, the flow past the plate remains laminar and only at about $0.05L$ away from the trailing edge of the sample, turbulent kinetic energy (Fig. \ref{fig:fluctuations}(a-c)), and Reynolds shear stress (Fig. \ref{fig:fluctuations}(d-f)) become visible and the location of the largest turbulent kinetic energy or the Reynolds shear stress is located at about $0.15L$ from the trailing edge for all cases. In addition, as it is seen from the vorticity distribution (Fig. \ref{fig:fluctuations}(g-i), normalized by a reference shear rate calculated based on $\dot{\gamma} = {U_{\infty}}/{\delta} = ({U_{\infty}}/{L}) \sqrt{{\rm Re}_L}$) and the accompanying streamlines in the wake of the samples, the extent of the separation bubble has a similar size across the various cases, with the bubble ending at around $0.1L-0.12L$ past the trailing edge of the sample.   

\begin{figure*}
    \centering
    \includegraphics[width = 1 \textwidth]{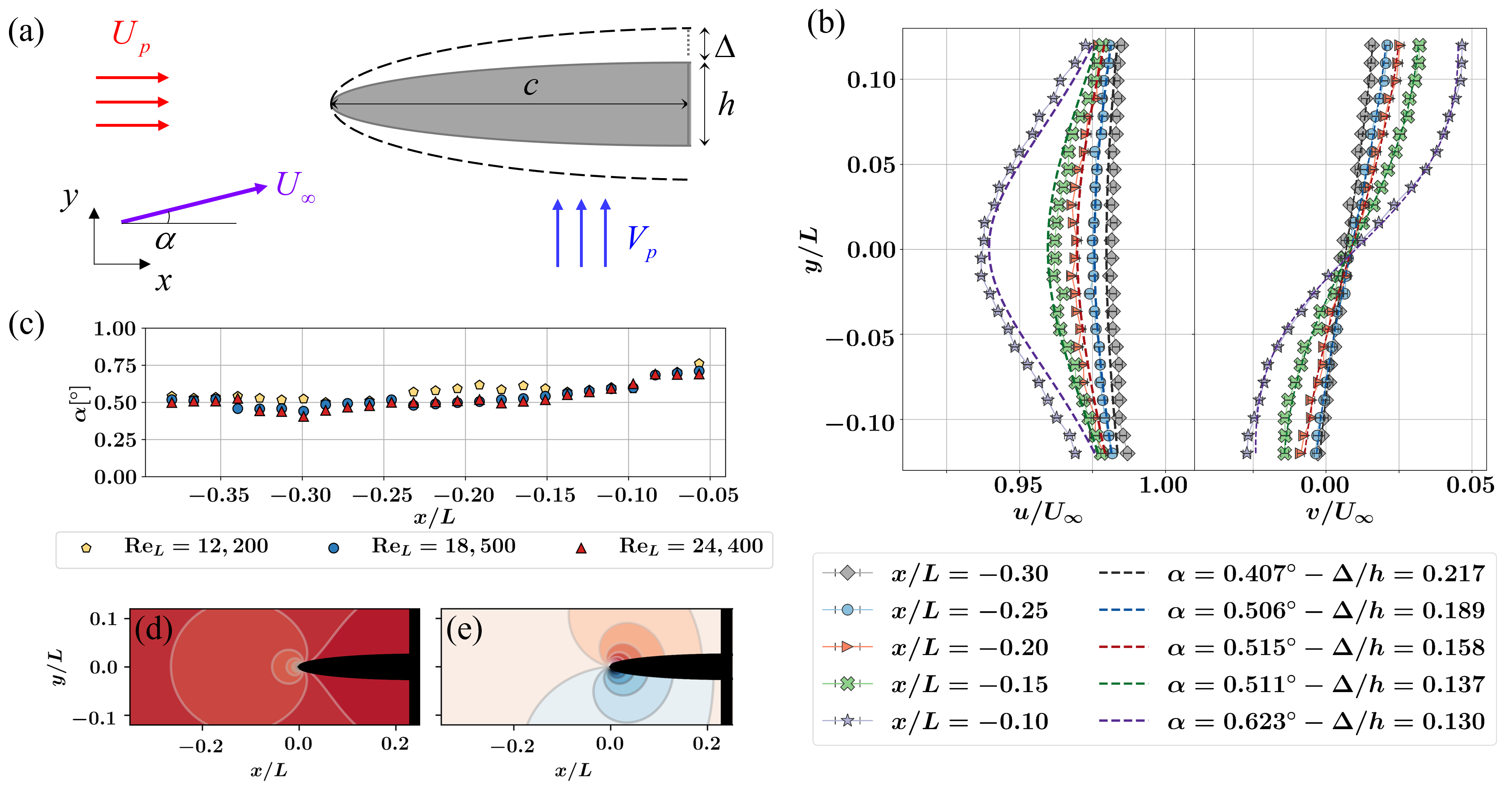}
    \caption{(a) Schematic of a half-ellipse with $c$ as the semi-major axis, and $h$ as the semi-minor axis, in a flow with free-stream velocity $U_{\infty}$ at an angle of attack $\alpha$, decomposed into $U_p$ in the $x$ direction, and $V_p$ in the $y$ direction where $\tan \alpha = V_p/U_p$. The effect of the boundary layer on the potential flow is represented by enlarging the semi-minor axis with $\Delta$ on both sides of the ellipse. (b) Velocity components in $x$ (left) and $y$ (right) at various $x$ locations as a function of $y$ for an example case with ${\rm Re}_L = 24,200$. Dashed lines are the potential flow fits with their respective $\alpha$ and $\Delta$. (c) The distribution of the calculated angle of attack from the fitting at various cross-sections as a function of $x/L$ for 3 cases at ${\rm Re}_L = 12,200$, ${\rm Re}_L = 18,500$, and ${\rm Re}_L = 24,200$. Far enough from the sample ($x/L<-0.1$), the angle of attack of all cases is on average $0.5^{\circ}$. Contours of theoretical velocity in (d) $x$ and (e) $y$ directions calculated using the potential flow theory, with average $\alpha = 0.5 ^{\circ}$, and $\Delta = 0.166 h$ from the fittings of part (b). Color limits (color bars) are the same as those in Fig. \ref{fig:abcd}.}
    \label{potential}
\end{figure*}

Lastly as seen in the contours of $v$ in Fig. \ref{fig:abcd}, the velocity distribution in the leading edge area is not symmetric about the $y=0$ line, in the same way as is expected from a symmetric sample aligned in the stream-wise direction. This hints at the possibility of a slight angle of attack in the sample placement with respect to the free-stream velocity. This angle is not visually detectable during the experiments, however, it can be deduced from the velocity fields.

\subsection{Estimation of the angle of attack}
\label{angle_attack}

To estimate the small angle of attack $\alpha$ in the experiments {($\phi+\chi$ in Fig. \mbox{\ref{fig:stitch_steps}})}, we use potential flow analysis. We assume the velocity field, far from the boundaries and in the vicinity of the leading edge can be described in terms of flow past an elliptical boundary. To factor in the angle of attack, we assume the free-stream velocity $U_{\infty}$ can be decomposed into $U_p$ in the $x$ direction, and $V_p$ in the $y$ directions with $\tan \alpha = V_p/U_p$ (see Fig. \ref{potential}(a)). Due to linearity, the complex potential ($w(z_p)$ where $z_p = x + i y$) of this flow is thus the superposition of complex potentials of the flow of $U_p$ in the horizontal direction ($w_{\parallel}$) and the flow of $V_p$ in the vertical direction ($w_{\perp}$) past an ellipse of the same orientation.

The closed-form solution of flow past an ellipse can be found using conformal mapping and the Zhukhovsky transformation between the complex variable $z_p = \zeta + {b^2}/{\zeta}$ and

\begin{equation}\label{zeta2}
    \zeta = \frac{1}{2}z_p - \frac{1}{2}\sqrt{z_p^2 - 4b^2}
\end{equation}

\noindent where an ellipse in $z_p$ plane of the form 

\begin{equation}\label{eq:elipse}
\dfrac{x^2}{\left(a+\dfrac{b^2}{a}\right)^2} + \dfrac{y^2}{\left(a-\dfrac{b^2}{a}\right)^2}=1
\end{equation}

\noindent is transformed into a circle of radius $a>b$ in $\zeta$ plane. 

{To account for near-wall viscous effects and the thickness of the boundary layer, we assume the body and the viscous boundary layer together have a fictitious boundary where the thickness of the sample as seen by the inviscid flow is larger than the sample boundaries. We represent this by assuming an ellipse with a semi-minor axis of $h/2+\Delta$ instead of $h/2$.} Therefore, using the dimensions of the sample (Fig. \ref{potential}(a)), and the above definition we have: 

\begin{equation}
        a + \dfrac{b^2}{a} = c \\
\end{equation}

\noindent and 

\begin{equation}
        a - \dfrac{b^2}{a} = \dfrac{h}{2} + \Delta.
\end{equation}

\noindent Thus, the full complex potential can be written as

\begin{equation}
    w = w_{\parallel} + w_{\perp} = U_p \left(\zeta + \dfrac{a^2}{\zeta}\right) -i V_p \left(\zeta + \dfrac{a^2}{\zeta}\right)
\end{equation}

\noindent with the velocities found using the chain rule

\begin{equation} \label{Com}
    \frac{dw}{dz_p} = u - iv = \dfrac{dw}{d\zeta}\dfrac{d\zeta}{dz_p}
\end{equation}

\noindent which are functions of the geometry of the elliptical leading edge, the angle of attack, and the average added thickness, $\Delta$. Now, given the geometry of the elliptical leading edge, and the measured velocity distribution, one can fit the above models to the measured velocities to find estimates for $\alpha$ and $\Delta$.

Fig. \ref{potential}(b) gives velocities $u$ (left) and $v$ (right) as a function of $y$ in a few $x$ locations (Symbols) in the flow field upstream of the leading edge for ${\rm Re}_L = 24,200$ ($x/L < 0$) and the dashed lines represent fitted results from potential flow solution. The error bars (which are very small) represent the $95\%$ confidence intervals. From the curve-fitting algorithm, one can see that on average the angle of attack is found to be around $\alpha \approx 0.5 ^{\circ}$ for this case (Figs. \ref{potential}(b) and \ref{potential}(c)). The same process has been repeated for multiple locations in the upstream of all three cases and as shown in Fig. \ref{potential}(c) all cases on average experience a similar angle of attack of $\alpha \approx 0.5 ^{\circ}$ using the fits for $x/L<-0.1$. Past $x/L = -0.1$ the near wall effects result in the potential flow model deviating from the measurements and thus they are not included in the calculation of the average angle of attack. It should be noted that these experiments were performed in one sitting where the sample was set up at the beginning of the day, the experiments repeated for the three velocities and then the sample is retrieved at the end and throughout the day the location of the sample is not adjusted or changed. Hence we do not expect the angle of attack of the sample to change between the three experiments as confirmed. This also confirms that the fixture was well secured in place and the flow (as expected) did not move or re-locate the position of the sample throughout the experiments.

Lastly, for comparison, contours of the normalized potential flow solutions, $u/U_{\infty}$ and $v/U_{\infty}$ are plotted in Figs. \ref{potential}(d-e) with the average value of $\alpha$ and $\Delta$ calculated from Fig. \ref{potential}(b) for the upstream and early elliptical portion of the leading edge and the results match with the experimental contours very well. Especially with the inclusion of the angle of attack, the $v/U_{\infty}$ contours show the line of $v=0$ which matches the experimental results very well and as expected is not symmetric about the line of $y=0$.

\subsection{Boundary layer} \label{BL}

To quantify the effect of the flow field on the wall, velocities in the boundary layers are used to calculate the distribution of the shear stress on both sides of the sample. This requires the calculation of the velocity gradient close to the wall. Numerical differentiation techniques with finite difference schemes only use a small portion of velocities adjacent to the wall (where the uncertainties can be larger than the rest of the profile) and are thus prone to cause large numerical errors. Therefore, to characterize the boundary layers and also calculate more accurate estimates of the velocity gradients at the wall, we choose to find the best possible functional fit to all the velocity measurements at each wall-normal direction, $\hat{n}$ (instead of finding a linear fit for a few points in the vicinity of the wall). {The functional form used for fitting is not chosen randomly and we employ the family of the Falkner-Skan solutions which are theoretical self-similar solutions to the boundary layer equation. }

\begin{figure*}
\centering
\includegraphics[width = 1\textwidth]{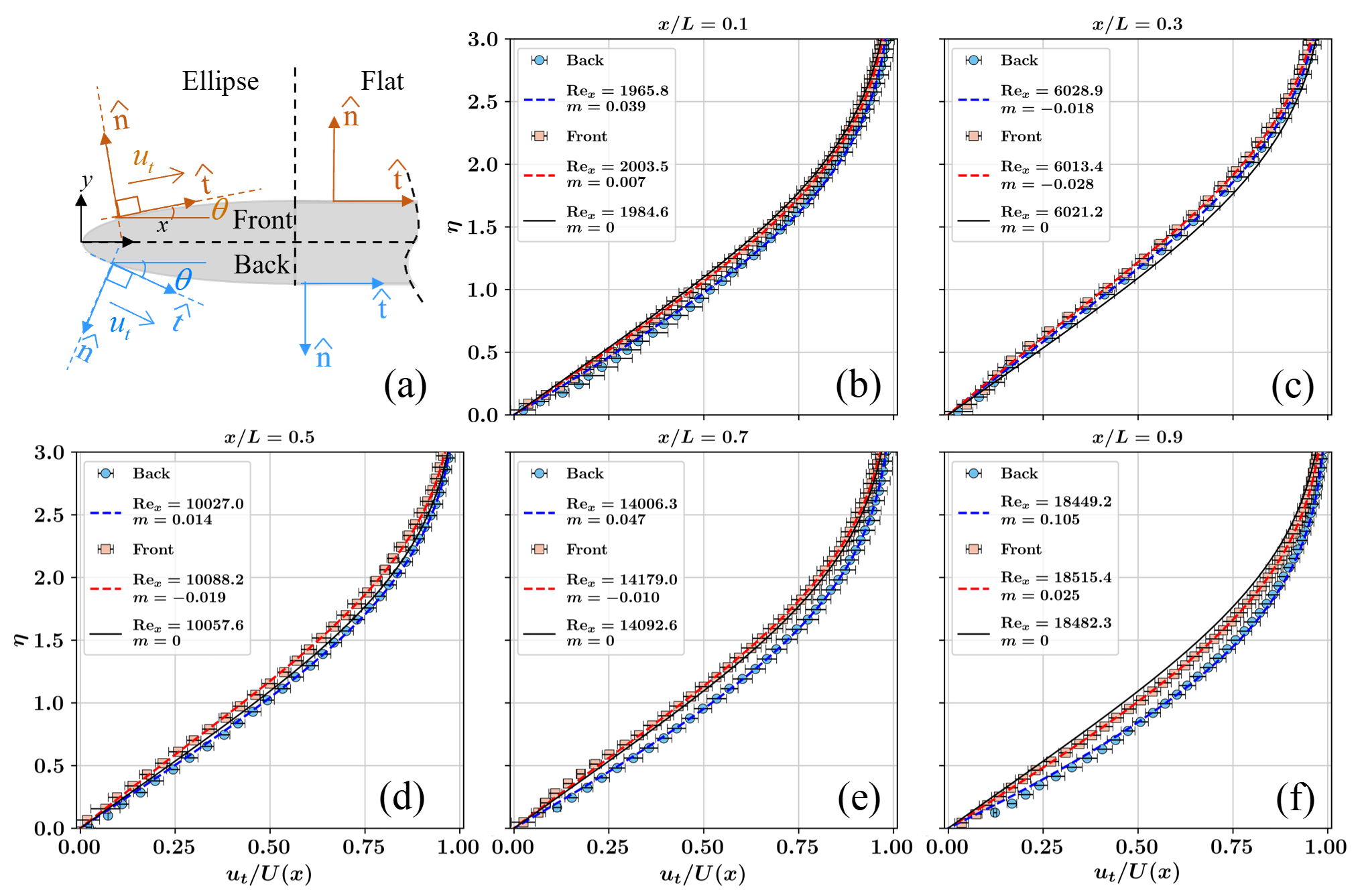}
\caption{(a) Schematic of the coordinate transformation from $(x,y)$ to $(t,n)$ (tangent and normal components) in the leading edge area. Past the leading edge $\hat{t}$ and $\hat{n}$ are the same as $x$ and $y$ coordinates. Velocity profiles on the Front (square) and Back (circle) of the sample as a function of $\eta$, at 5 different positions along the length, (b) $x/L = 0.1$, (c) $x/L = 0.3$, (d) $x/L = 0.5$, (e) $x/L = 0.7$, and (f) $x/L = 0.9$, with the corresponding Falkner-Skan fits for experiments performed at ${\rm Re}_L = 18,500$. The Blasius solution is shown with a solid black line on all the plots. The error bars represent the 95\% confidence intervals. }
\label{fig:vel_placeholder.png}
\end{figure*}

{From Falkner-Skan theory \mbox{\citep{Schlichting2014boundary}}, the boundary layer in the tangential direction, $\hat{t}$, to the wall (see Fig. \mbox{\ref{fig:vel_placeholder.png}(a))} is defined using the local Reynolds number ${\rm Re}_{x}$ and a parameter $m$, covering solutions for boundary layer profiles with positive, $m>0$, that are more attached than the Blasius solution ($m=0$) and those that are more separated from the wall at negative values of $m$ within $0>m>-0.0904$ where $-0.0904$ is the lowest possible value mathematically.} Thus, $u_t$ is defined as a function of the local Reynolds number ${\rm Re}_{x} = {\rho x U(x)}/{\mu}$, $n$ and $m$

\begin{equation}\label{eq:u}
     u_t  = \mathcal{G}({\rm Re}_{x},n;m)
\end{equation}

\noindent and in the self-similar form of the Falkner-Skan theory, the velocity is written as

\begin{equation}\label{eq:FSu1}
    \frac{u_t}{U} = \mathcal{F}'(\eta)
\end{equation}

\noindent with $\eta$ defined as

\begin{equation}\label{eq:eta}
    \eta = \frac{n}{x} \sqrt{{\rm Re}_x \left(\dfrac{m+1}{2}\right)}.
\end{equation}

Knowing the spatial location $(x,n)$ and the velocity distribution $u_{t}(x,n)$ from the experimental data, we employ a least-square fitting algorithm to find the best $m$ for the velocity profiles at different $x$ locations along the length of the sample on either side. As the input to the curve-fitting, we use the measured velocity profiles $u_t$ from all 50 image pairs of the step and their corresponding wall-normal, $n$, locations from the wall up to the location where $u_t$ is the maximum in the given normal direction. We consider this maximum location as the edge of the boundary layer where the inner solution (boundary layer solution) meets the outer solution from the inviscid flow \citep{kundu2015fluid}. As a result of this $U(x)$, used in the definition of the local Reynolds number ${\rm Re}_x$, is not a constant and as shown earlier in Sec. \ref{Results} \ref{contours} is nearly always higher than $U_{\infty}$. Therefore, the local Reynolds number ${\rm Re}_x = \rho U(x) x/\mu$ is always larger than $\rho U_{\infty} x/\mu$. {It should be noted that one can choose any other family of functions for this purpose, however, the ability of the Falkner-Skan theory to capture a wide range of velocity profiles more attached or detached compared to the Blasius solution (i.e. non-zero pressure gradients) makes this family a more attractive choice here.}

The family of Falkner-Skan functions is implemented in the form of an ODE solver with the CasADi package \citep{Andersson2019, FS} in Python. In case of the presence of outliers within the data, the curve-fitting procedure is augmented with a RANSAC (RANdom SAmple Consensus) \citep{fischler1981random} algorithm, and only the experimental data identified as inliers are used in the final curve-fitting procedure. The inlier threshold in the algorithm is set to ensure that more than 95\% of the data are considered inliers.

In the flat part of the samples (past the elliptic leading edge), the wall-normal direction is the same as the $y$ direction ($n = \lvert y \lvert - h/2$), Fig. \ref{fig:vel_placeholder.png}(a)), which is not the case within the elliptical leading edge. In this region, we find the normal to the wall, $n$, at every $x$ location from the equation of the corresponding ellipse and use two-dimensional interpolation to find the distribution of the velocity in the local tangential direction, $u_t(n) = u \cos \theta + v \sin \theta$ where $\theta$ is the angle of the local tangent at the wall (Fig. \ref{fig:vel_placeholder.png}(a)) and use these values and the corresponding calculated normals in the curve-fitting procedure.

{The Falkner-Skan family of solutions is generated based on the theoretical assumption of flow past a wedge with a local free-stream speed defined as $U(x) \propto x^m$, where $m$ is constant throughout the $x$ direction. However, in this work, we assume that $m$ is just a mathematical parameter and only use the family of Falkner-Skan solutions as a set of mathematical functions available to describe the shape of the boundary layers. The purpose here is not to find the closest Falkner-Skan fit for the entire flow but to find the local best fits to the experimental velocity profiles at each $x$ location for more accurate post-processing steps, specifically calculations of the shear stress distribution.}

Velocity profiles in the boundary layer at 5 different locations along the length of the sample operated at the global ${\rm Re}_L = 18,500$ (calculated with $U_{\infty}$), on the Front and Back sides, are presented in Figs. \ref{fig:vel_placeholder.png}(b-f) as a function of the similarity variable $\eta$ (Eq. \ref{eq:eta}). 
The first thing to note in all these figures is that even though the sample is symmetric, the small angle of attack in the experiments leads the velocity profiles on either side of the samples to take different shapes at the same location as shown by the different fitted values of $m$ for each side. At $x/L=0.1$, which is within the elliptic leading edge, the velocity profile on the Back has an $m>0$ while the Front profile is nearly close to a Blasius profile with $m$ slightly larger than 0. Then moving to $x/L = 0.3$, the profiles on either side look more detached than the Blasius solution with negative $m$ values. However, at $x/L = 0.5$ the profile on the Back side moves to a more attached form with $m = 0.014$ increasing to $m = 0.047$ and $m = 0.105$ at $x/L= 0.7$ and $x/L = 0.9$ respectively. This is while at $x/L = 0.5$ and $x/L = 0.7 $ the velocity profiles on the Front still maintain a negative $m$ and only at $x/L = 0.9$ the velocity profile becomes more attached to the surface with $m = 0.025$. As is clear from all cases, the $m$ values for the velocity profiles on the Back are always larger than those on the Front (flow is more attached on the Back than the Front).

\begin{figure}
\centering
\includegraphics[width = 0.5\textwidth]{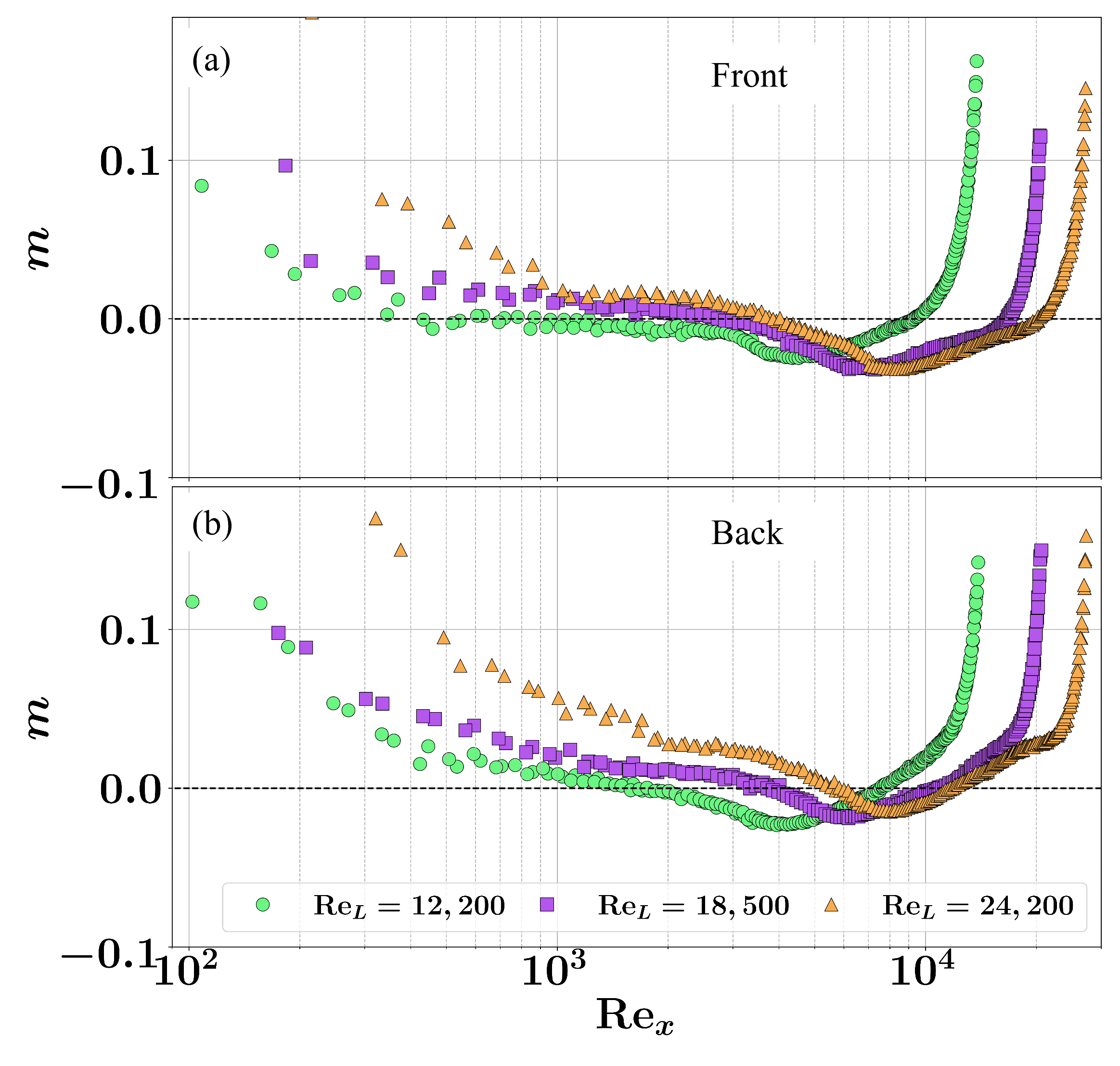}
\caption{Distribution of the parameter $m$ for boundary layers on the (a) Front and (b) Back sides of the samples operated at ${\rm Re}_L = 12,200$, ${\rm Re}_L = 18,500$, and ${\rm Re}_L = 24,200$ as a function of the local Reynolds number. Note that as discussed in the text, the local Reynolds number at $x=L$ is larger than ${\rm Re}_L$. {Note that no gaps or jumps are visible in the plots, confirming the ability of the consecutive imaging to capture nearly identical results in the overlapping steps.} }
\label{fig:m_Rex.png}
\end{figure}

Instead of presenting all the velocity profiles along the length, the distribution of the parameter $m$ as a function of the local Reynolds number ${\rm Re}_x = \rho U(x) x/\mu$ is illustrated in Fig. \ref{fig:m_Rex.png} for the Front and Back sides of the sample for all the three experimental cases. Note that, here, the curve-fitting process is performed with the 50 instantaneous velocity measurements of each separate step and repeated for all steps of the experiment (excluding the ones capturing the space before the leading edge and after the trailing edge of the sample) and Fig. \mbox{\ref{fig:m_Rex.png}} includes the $m$ parameters calculated for all the steps overlaid on each other. The distribution of $m$ in the different cases is qualitatively similar and only stretched out when plotted as a function of the local Reynolds number ${\rm Re}_x$. If plotted as a function of the normalized horizontal location $x/L$, the distribution of the $m$ for the different cases are similar with a slight bit of upward shift moving toward higher ${\rm Re}_L$ cases. Hence one observes a stronger dependence on the length of the sample than on the local value of the Reynolds number.

The second point to highlight is that throughout the length, $m$ for all the cases starts at $m>0$ which corresponds to a region of favorable pressure gradient, and then crosses over to $m<0$ where the flow then experiences an adverse pressure gradient, but this does not last all the way and toward the trailing edge again the flow experiences a favorable pressure gradient and $m$ crosses over to $m>0$. Thus, even though $75\%$ of the length of the sample consists of a flat plate (on either side), the local velocity in the boundary layers does not fully follow the Blasius boundary layer at a zero pressure gradient and the resulting $m$ parameter shows a distribution along the length where the profiles are initially more attached and then more detached and later more attached to the wall compared to the Blasius solution. 

{Lastly, we should note that the largest boundary layer thickness captured on either side of the sample in terms of $\delta_{99} = 0.99 U(x)$ is found to be 3.71 mm on the Front and Back at ${\rm Re}_L = 12,200$, 3.23 mm on the Front, and 3.07 mm on the Back for ${\rm Re}_L = 18,500$, and 2.74 for the Front, and 2.67 for the Back at ${\rm Re}_L = 24,200$. Thus, again we confirm that a 1D camera sweep and the field of view of 25 mm in the normal direction is enough to capture the details of the boundary layers.}

\subsection{Shear stress distribution} \label{shear}

Knowing the mathematical form of the Falkner-Skan solutions as well the distribution of the $m$ parameter, the local shear stress distribution along each side of the plate can be calculated using $m$ and derivatives of Eq. \eqref{eq:u} with respect to $n$ as

\begin{equation}\label{eq:tau}
   \tau   (x) = \left. \dfrac{\partial  u_t }{\partial n} \right \vert_{n=0} = \left. \left(\frac{m+1}{2}\right)^{0.5}\frac{\rho U(x)^2}{\sqrt{{\rm Re}_x}}\mathcal{F}''\right \vert_{\eta=0}
\end{equation}

\noindent and the skin friction coefficient is then determined by 

\begin{equation}\label{eq:Cf}
    C_f(x) = \frac{  \tau(x)  }{\frac{1}{2}\rho U(x)^2}.
\end{equation}

\begin{figure}
\centering
\includegraphics[width = 0.5\textwidth]{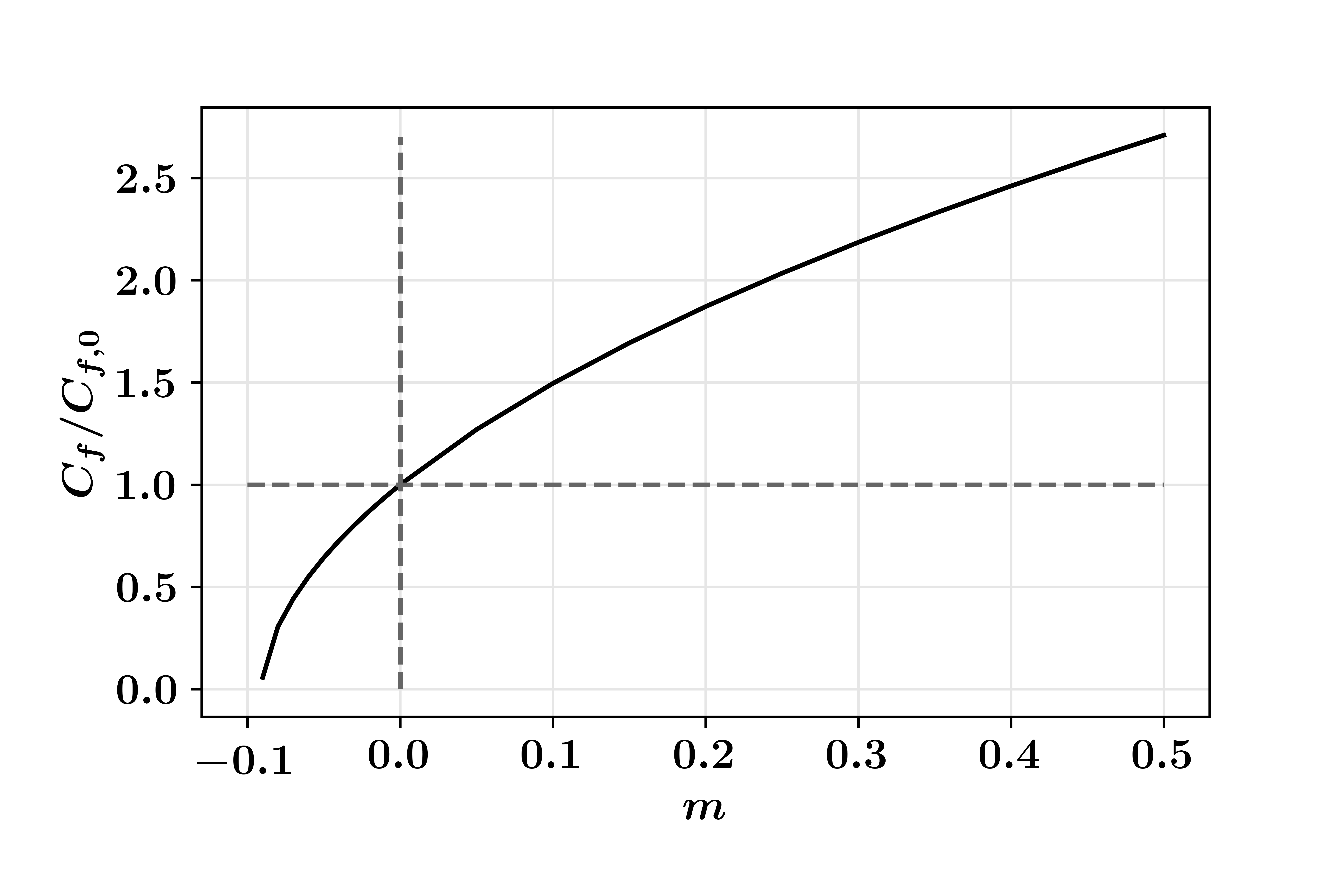}
\caption{The ratio of the skin friction coefficient of the Falkner-Skan family of boundary layers normalized by the skin friction coefficient of the Blasius boundary layer ($m=0$) as a function of $m$.}
\label{fig:FS_compare.png}
\end{figure}

The variation in $m$ as seen in Fig. \ref{fig:m_Rex.png} results in a shear stress distribution different from that of the Blasius boundary layer. Using Eqs. \eqref{eq:tau} and \eqref{eq:Cf} one can see that as shown in Fig. \ref{fig:FS_compare.png} for $m<0$ in cases where the boundary layer is less attached to the wall than the Blasius solution ($m=0$) the shear stress experienced is less than the $m=0$ case and for more attached boundary layers with $m>0$, the shear stress can reach as much as 1.5 times the shear stress from the Blasius solution at $m=0.1$. 

\begin{figure*}
\centering
\includegraphics[width = 1\textwidth]{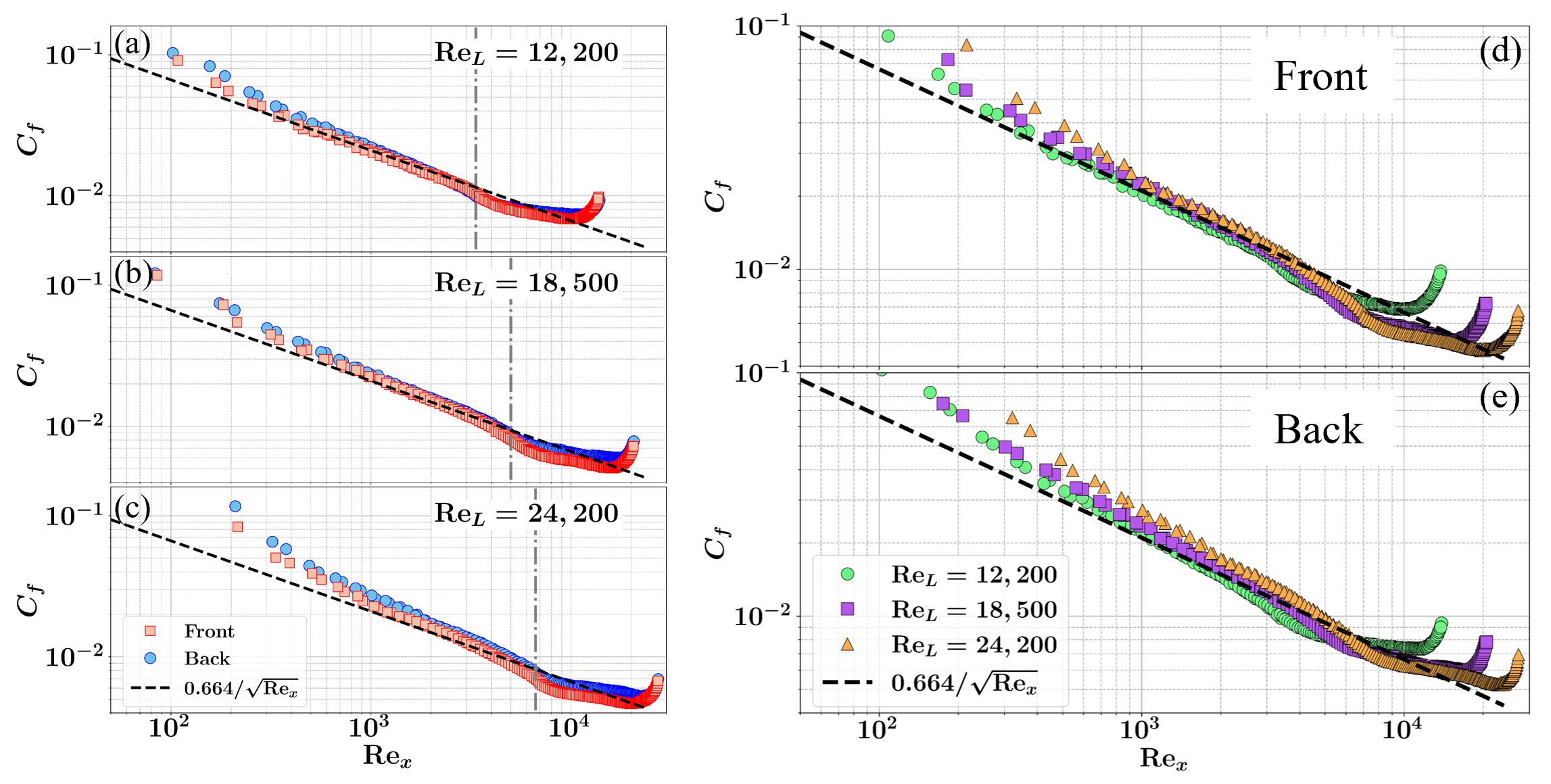}
\caption{Distribution of the skin friction coefficient on the Front (square) and Back (circle) of the sample at (a) ${\rm Re}_L = 12,200$, (b) ${\rm Re}_L = 18,500$, and ${\rm Re}_L = 24,200$. The dash-dotted lines denote the location of the end of the elliptical leading edge and the black dashed lines are the theoretical shear stress calculated from the first-order boundary layer theory (Blasius Solution). All the skin friction results are also separated for (d) Front and (e) Back sides, which are overlaid on each other. Local Reynolds number is calculated based on the maximum velocity $U(x)$ along the corresponding normal which is also used in the curve-fitting process.}
\label{fig:Cf_Re.png}
\end{figure*}

The variations in the skin friction coefficient on the Front and Back of the sample tested at three different Reynolds numbers are shown in Fig. \ref{fig:Cf_Re.png}(a-c) as a function of the local Reynolds number. In all cases, the skin friction coefficient experienced on the Front side is slightly lower than that of the Back which is also visible in the distribution of the $m$ where at the same location $m$ on the Front side is slightly lower than the $m$ on the Back. In the case with ${\rm Re}_L = 12,200$ (Fig. \ref{fig:Cf_Re.png}(a)), the skin friction coefficient on the Front and Back sides of the sample are the closest, and as the global ${\rm Re}_L$ is increased by increasing the $U_{\infty}$ the difference between the two sides become more visible (Figs. \ref{fig:Cf_Re.png}(b) and \ref{fig:Cf_Re.png}(c)). 

While in the leading edge area, the skin friction coefficient decreases at a rate faster than ${\rm Re}_x^{-0.5}$, past the elliptic leading edge and in the flat part of the sample, the skin friction coefficient experiences a slower rate of change compared with the ${\rm Re}_x^{-0.5}$. Also, close to the trailing edge, the skin friction coefficient reverses its course and goes through an increasing trend which has also been predicted in second-order models of the boundary layer over a flat finite plate \citep{dennis1966steady}.

Similar to $m$, the qualitative trends in the shear stress results, for both the Front and Back sides, show a strong dependence on the length of the sample, more than the local Reynolds number as is seen in Figs. \ref{fig:Cf_Re.png}(d) and \ref{fig:Cf_Re.png}(e). In addition, early on, within the leading edge area, an increase in the ${\rm Re}_L$ results in an increase in the skin friction recorded at similar local Reynolds numbers, however, as we move to the flat area, the skin frictions cross over each other where for the rest of the length, the case with the lowest ${\rm Re}_L$ experiences the largest skin friction coefficient among all.

\subsection{Forces} \label{all_force}

\begin{figure*}
\centering
\includegraphics[width = 1\textwidth]{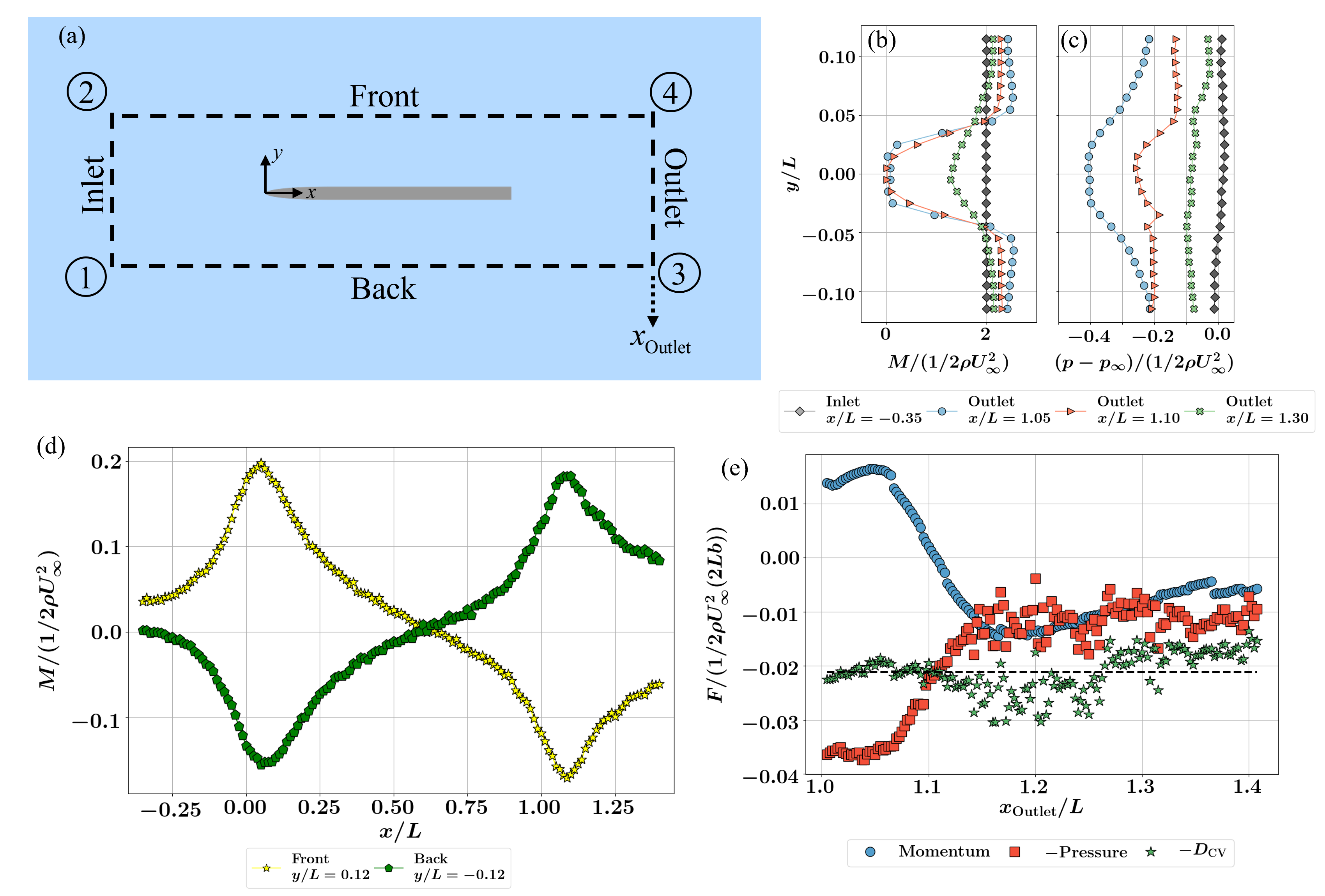}
\caption{(a) Schematic of the rectangular control volume used here, the Inlet is located at $x/L = -0.35$, the Front and Back boundaries are located at $y/L = \pm 0.12$, and the Outlet boundary is moved from $x/L = 1.00$ up to $x/L = 1.42$. Distribution of the (b) momentum terms (M, Eq. \eqref{CV1}) and (c) the pressure at the Inlet and a few Outlet positions. (d) Distribution of the Momentum (M) on the Front and Back boundaries. (e) The calculated integrals of the Momentum terms (blue circles), and negative of the integral of pressure (red squares), and the negative of the resulting total drag (green stars) as a function of the location of the Outlet boundary. All calculated force integrals are non-dimensionalized in the form of a drag coefficient. The dashed black line is the mean of the negative of the total drag force values. All the results are for the case of ${\rm Re}_L = 12,200$.}
\label{fig:CV}
\end{figure*}

Drag force is the force component experienced by the sample in the streamwise direction. Here due to the small angle of attack, the drag force on the sample is written as 

\begin{equation}
    D = F_x \ \cos \alpha + F_y \sin \alpha 
\end{equation}

\noindent where $\cos \alpha = 0.999962$ and $\sin \alpha = 0.0087$. So, we assume that $D \approx F_x$. 

Total drag experienced by a finite-thickness sample is decomposed into contributions from the viscous effect in the boundary layer and the pressure distribution around the sample. The vicious part of the drag force is calculated from the integral of the shear stress distribution (Fig. \ref{fig:Cf_Re.png}) on either side of the sample

\begin{equation}
\begin{split}
         &D_{\rm viscous} = \int_0^L \pmb{\tau} \cdot \mathbf{n} dA = \\
         &b \left(\int_0^L \tau_{_{\rm Front}}(x) dx + \int_0^L \tau_{_{\rm Back}}(x) dx \right)
\end{split}
\end{equation}

\noindent where it is assumed that shear stress distribution measured in the mid-section is constant throughout the span of the sample (i.e. $\tau(x,z) \approx \tau(x)$). In the elliptic leading edge of the sample, the element of the integral is $\tau(x) \cos \theta dt$ which is the same as $\tau(x) dx$ (see Fig. \ref{fig:vel_placeholder.png}(a)), and thus the latter form is applicable to the entire length of the sample.

While the tested sample is slender ($h/L = 0.05$), the finite thickness of the sample has non-negligible effects on the distribution of pressure in the flow and as a result, the total drag force includes a contribution from pressure, also known as form drag. However, the effect of the pressure distribution cannot be found independently and is found in a cumulative manner with the viscous drag using a rectangular control volume (Fig. \ref{fig:CV}(a))

\begin{equation}
\begin{split}
         -D_{\rm CV} \ &+  \overbrace{\int_{S_{_{\rm Inlet}}} pdA - \int_{S_{_{\rm Outlet}}}pdA}^{\rm Pressure \ Force \ on \ Boundaries} = \\  &\overbrace{\sum_{i} \int_{S_i} \underbrace{\rho \tilde{u} (\tilde{\mathbf{u}} \cdot \mathbf{n})}_{M} dA_i }^{\rm Momentum}
    \label{CV1}
\end{split}
\end{equation}

\noindent where $i \in [ {\rm Inlet}, \ {\rm Outlet}, \ {\rm Front}, \ {\rm Back}]$, and the total of the reaction force in the $x$ direction (which is, according to Newton's third law, the negative of the total force applied on the sample, i.e. $-D_{\rm CV}$) and the pressure forces equals the variations in the momentum crossing the boundaries of the control volume. {($D_{\rm CV}$ is the total drag force experienced by the sample found using the control volume analysis.)} It should be noted that the momentum terms in Eq. \eqref{CV1} also include the effect of the pressure distribution. Therefore, $D_{\rm form} = D_{\rm CV}- D_{\rm viscous}$.

In steady-state form, Eq. \eqref{CV1} can be expanded to include the effect of the Reynolds stresses in the flow field and written in terms of a Reynolds-averaged integral momentum (RAIM) conservation equation \citep{Ferreira_2021} in the form of

\begin{equation}
\begin{split}
    &-D_{\rm CV} \ + b{\int (p_{_{\rm Inlet}} - p_{_{\rm Outlet}}}) dy  = \\ & \rho b \left( \int_{\rm Outlet} ({u} {u} + \overline{u'u'}) dy - \int_{\rm Inlet} ({u} {u} + \overline{u'u'}) dy \right) + \\
    & \rho b \left( \int_{\rm Front} ({u} {v} + \overline{u'v'}) dx -  \int_{\rm Back} ({u} {v} + \overline{u'v'}) dx \right)
\end{split}
\end{equation}

\noindent with $\tilde{u} = u + u'$ and $\tilde{v} = {v} + v'$ where ${u}$ and $v$ are the means of the velocity components and $u'$ and $v'$ are the fluctuation terms. This formulation thus includes the effect of the Reynolds stresses on the total force calculations, and even though small they have all been incorporated in the current analysis.

In an ideal setup, where the experiments are performed in unbounded flows with access to far-field information farther than multiple body lengths away, one can choose the control volume boundaries far enough where the local pressure and velocity at the boundaries are back to $U_{\infty}$ and $p_{\infty}$. There, only the momentum components of the control volume would be sufficient for finding the reaction force as commonly discussed in fluid mechanics textbooks \citep{kundu2015fluid, batchelor1967introduction}. In such a case, the pressure difference across the inlet and outlet faces will be zero. However, with the physical limits available in the experimental setup, the local pressure especially downstream of the flow does not fully recover to $p_{\infty}$, and thus we do not neglect this term as has been done in previously reported investigations \mbox{\cite{michalek2022influence, terra2016drag}}. 

Hence, we use the two-dimensional Reynolds-averaged Navier-Stokes equations and directional integration to find the pressure distribution \cite{Charonko_2010} 
 on the boundaries of the control volume where local pressure along a horizontal ($y$ constant) line and vertical line ($x$ constant) can be calculated using 

\begin{equation}
\begin{split}
         & p(x)-p(x_{\rm ref})  = \int_{x_{\rm ref}}^{x} {-\rho \left(u \dfrac{\partial u}{ \partial x} + v \dfrac{\partial u}{\partial y} \right)}  \\ &+ \mu \left(\dfrac{\partial^2 u}{\partial x^2} + \dfrac{\partial^2 u}{\partial y^2} \right) - \rho \left( \dfrac{\partial \overline {u'u'}}{\partial x} + \dfrac{\partial \overline {u'v'}}{\partial y}\right) dx
    \label{p_x}
    \end{split}
\end{equation}

\noindent and

\begin{equation}
\begin{split}
         & p(y) - p(y_{\rm ref}) =  \int_{y_{\rm ref}}^{y}{-\rho \left(u \dfrac{\partial v}{ \partial x} + v \dfrac{\partial v}{\partial y} \right)}  \\ &+ \mu \left(\dfrac{\partial^2 v}{\partial x^2} + \dfrac{\partial^2 v}{\partial y^2} \right) - \rho \left( \dfrac{\partial \overline {u'v'}}{\partial x} + \dfrac{\partial \overline {v'v'}}{\partial y}\right) dy
    \label{p_y}
\end{split}
\end{equation}

\noindent respectively.

Here, assuming points {\large \textcircled{\small 1}}  and {\large \textcircled{\small 2}} (Fig. \ref{fig:CV}(a)) are at $p_{\infty}$ and we use equation \eqref{p_y} in both positive ({\large \textcircled{\small 1}}  $\rightarrow$ {\large \textcircled{\small 2}}) and negative ({\large \textcircled{\small 2}}  $\rightarrow$ {\large \textcircled{\small 1}}) directions to calculate $p^{+}(y)$ and $p^{-}(y)$ on the Inlet and use an average of the two ($p_{_{\rm Inlet}}(y) = (p^{+}(y) + p^{-}(y))/2$) for the pressure distribution at $x/L = -0.35$ (Fig. \ref{fig:CV}(c)) which is nearly constant at $p_{\infty}$. From there, we integrate Eq. \eqref{p_x} in the positive direction from {\large \textcircled{\small 1}}  $\rightarrow$ {\large \textcircled{\small 3}} and {\large \textcircled{\small 2}}  $\rightarrow$ {\large \textcircled{\small 4}} on the Front and Back to calculate the pressure distribution $p(x)$ on either boundary. For boundaries at $\lvert y \rvert /L>=0.08$ the pressure distribution calculated this way is identical to the pressure calculated from the Bernoulli equation (flow response is inviscid). Below that as one gets closer to the wall, the viscous effects result in larger and larger deviations from that of the Bernoulli equation. Now, knowing the pressure at points {\large \textcircled{\small 3}} and {\large \textcircled{\small 4}} (Fig. \ref{fig:CV}(a)), then we once again use Eq. \eqref{p_y} and integrate it in both positive ({\large \textcircled{\small 3}}  $\rightarrow$ {\large \textcircled{\small 4}}) and negative ({\large \textcircled{\small 4}}  $\rightarrow$ {\large \textcircled{\small 3}}) directions and use an average of the two ($p_{_{\rm Outlet}}(y) = (p^{+}(y) + p^{-}(y))/2$) for the pressure distribution on the Outlet. A few examples are shown in Fig. \ref{fig:CV}(c) for pressure distribution at different Outlet positions ($x_{_{\rm Outlet}})$.

The size of the control volume should not matter in the calculation of the $D_{\rm CV}$ as long as all the forces applied to the control volume are accounted for. Thus, we choose the boundary of the Front and Back position to be far away from the boundary layer so that there are no shear stresses applied on those boundaries. Here we present the results for the Front and Back boundaries fixed at $\lvert y \rvert /L = 0.12$. Similarly, we keep the Inlet far from the leading edge where the inlet velocity is nearly constant at $x/L = -0.35$ (Fig. \ref{fig:CV}(b)) and move the outlet boundary from the trailing edge of the sample up to $x/L = 1.42$. 

The choice of the location of the Inlet allows for the distribution of the pressure and momentum terms ($M$) to be constant along this boundary as shown in Figs. \ref{fig:CV}(b) and \ref{fig:CV}(c) (grey diamonds). However, along the Outlet, the momentum distribution ($\rho(u^2 + \overline{u'u'})$) follows the form of velocity deficits expected from a wake (Fig. \ref{fig:CV}(b)). As a result, at the trailing edge of the sample, the pressure is also lower and raises as one moves further away from the trailing edge. This pressure distribution is visible very close to the trailing edge and as one moves away and the wake diffuses away, the pressure reaches to near constant at $x/L \approx 1.3$ but does not fully recover to $p_{\infty}$ (the inviscid velocity also stays larger than $U_{\infty}$ and does not fully recover within the region of study). 

On the Front and Back the distributions of the momentum ($\rho (uv + \overline{u'v'})$, see Fig. \ref{fig:CV}(d)) follow very similar trends (just with opposite signs) with Front(Back) experiencing an increase(decrease) due to flow being pushed away from the sample at the leading edge area and then a decrease(increase) past the trailing edge due to the flow being pulled toward the center-line. However, the small angle of attack results in the two momentum distributions being symmetric about a non-zero value where in the leading edge area the momentum term is positive on the $y/L = 0.12$ line while it is zero on the $y/L = -0.12$.

Within a control volume, in addition to momentum, mass also needs to be conserved. However a quick survey of the flow rates in and out of the 4 boundaries of the control volume in Fig. \ref{fig:CV}(a) shows that the total mass flow in and out of the control volume is not strictly zero and there is potential leakage in the up/down ward directions due to the three-dimensional nature of the problem. Thus, to account for this, we assume that the control volume is 3D where the fifth and sixth boundaries are at $z=0$ (Bottom) and $z=b$ (Top) locations. Even with access to 2D velocity distribution, we can use the continuity equation to find the total mass flux across both the Top and Bottom boundaries as

\begin{equation}
    \dot{m} = -\sum_i \rho \int_{S_{i}} \mathbf{\tilde{u}}.\mathbf{n} dA
\end{equation}

\noindent where $i \in [ {\rm Inlet}, \ {\rm Outlet}, \ {\rm Front}, \ {\rm Back}]$. Then one can estimate the momentum flux through these two surfaces as

\begin{equation}
    {\cal M}_{_{\rm Top + \rm Bottom}} = \dot{m} \overline{u} 
\end{equation}

\noindent with $\overline{u}$ as an average of the $u$ along all other four boundaries 

\begin{equation}
    \overline{u} = \dfrac{1}{4}\left({\sum_i} \dfrac{\int u d{\ell}}{\int d{\ell}} \right)
\end{equation}

\noindent Therefore Eq. \eqref{CV1} is updated to 

\begin{equation}
\begin{split}
         -D_{\rm CV} \ &+ \overbrace{\int_{S_{_{\rm Inlet}}}pdA - \int_{S_{_{\rm Outlet}}}pdA}^{\rm Pressure \ Force \ on \ Boundaries} = \\ & \overbrace{\sum_{i} \int_{S_i} \underbrace{ \rho \tilde{u} (\tilde{\mathbf{u}} \cdot \mathbf{n})}_{M} dA_i + {\cal M}_{_{\rm Top + \rm Bottom}} }^{\rm Momentum} .
    \label{CV}
\end{split}
\end{equation}

\begin{figure}
\centering
\includegraphics[width =0.4\textwidth]{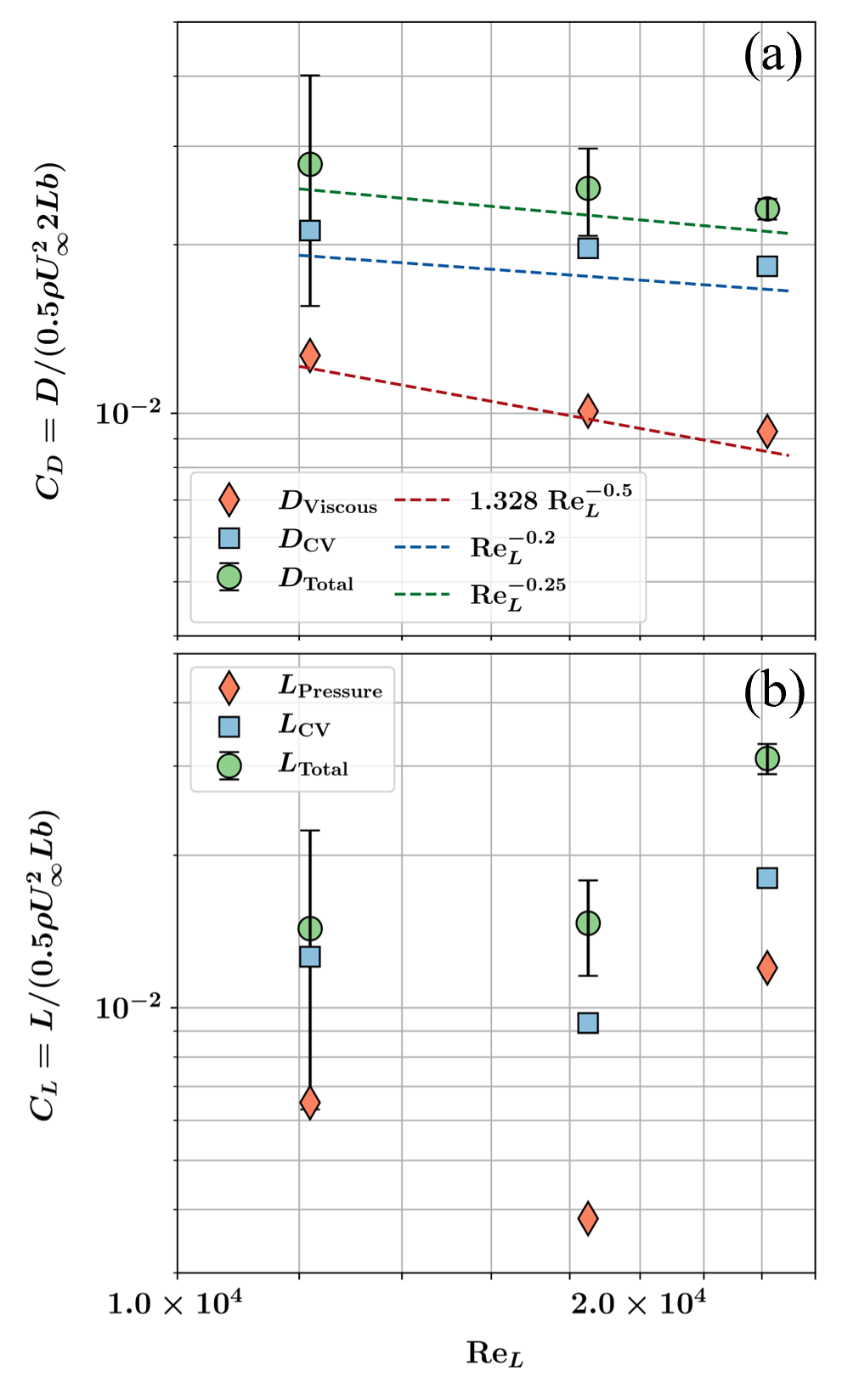}
\caption{(a) Drag coefficient and (b) Lift coefficient for the sample at three different Reynolds numbers and decomposed in terms of various effects. Drag is decomposed in terms of the viscous drag, drag calculated using the control volume, and the total drag measured using the load cell. {Lines ${\rm Re}_L^{-0.2}$ and ${\rm Re}_L^{-0.35}$ are plotted as visual guides.} Lift is decomposed in terms of the pressure difference between the Front and Back, lift found from control volume analysis, and lift measured with the load cell. Error bars represent the 95\% confidence intervals of the load cell measurements and they decrease as the magnitude of the measured load increases. }
\label{fig:Cd_ReL.png}
\end{figure}

Putting all the terms together, we can find the total of the pressure forces, total of the momentum contributions, and $D_{CV}$ experienced by the sample as a function of different Outlet positions ($x_{\rm Outlet}$) as presented in Fig. \ref{fig:CV}(e). As it can be seen, very close to the trailing edge ($1<x/L<1.1$), the negative of the total pressure forces and the total integral of the Momentum terms are not constant and they go through a variation with a reduction in the magnitude of both, where the sum of the two results in a constant force experienced by any control volume as a function of the location of the Outlet planes within ($1<x_{\rm Outlet}/L<1.12$). Past this point, the variations in the (negative) of the pressure forces and the integral of the Momentum contribution subsidies. However, the scatter in the Pressure term increases due to the increase in numerical errors in the calculation of the derivatives of $v$ in the wake area where the magnitude of the velocity decreases as $x$ is increased. In addition, both the Momentum and Pressure terms look like they are oscillating about a constant mean force (dashed black line in Fig. \ref{fig:CV}(e)).

Also, as was shown in Fig. \ref{fig:fluctuations}, from $x/L \approx 1.05$ the fluctuation terms start to gain strength and the flow slowly becomes more turbulent with the location of the largest turbulent kinetic energy and Reynolds shear stress being around $x/L \approx 1.15$. The appearance and enhancement of the turbulence statistics also coincide with the region of this oscillatory behavior in the force calculation and require further investigation as to its nature. In addition, $x/L \approx 1.1-1.12$ is also the ending point of the separation bubble behind the sample, and the vortex shedding behind this point could possibly affect the results beyond $x/L \approx 1.12$ \citep{chopra2019drag}, which needs to be further investigated using a higher frequency or fully time-resolved measurement.

Lastly, the total drag force is also affected by the three-dimensional nature of the sample and the flow which is not fully captured with a 2D-2C PIV measurement. This total load can be found from the load-cell measurements that we conducted simultaneous to the PIV measurements and a summary of all the forces is presented in terms of drag coefficients, $C_D = D/(1/2 \rho U_{\infty}^2 (2Lb))$ in Fig. \ref{fig:Cd_ReL.png}(a). As shown in the figure, each level of the analysis presented here allows us to capture contributions of the different phenomena on the drag force from the viscous and pressure parts, to the effects of the finite 3D nature of the sample. As expected, $D_{\rm viscous} < D_{\rm CV} < D_{\rm total}$. On average the viscous part of the drag is about $40-45\%$ and the form drag is about $30-38\%$ of the total drag, leaving about $25-30\%$ to the 3D effects of the sample. 

Overall, due to the slender nature of the sample, the viscous drag takes the largest portion of the total drag, however, the pressure drag is not negligible (as is usually the assumption when dealing with flow past a flat plate) and the contribution rises as the Reynolds number is increased. {While the drag coefficient due to viscous effects visually follows a ${\rm Re}_L^{-0.5}$ trend, the drag coefficient from the control volume follows a lower rate of change. As the Reynolds number is increased there is a lower decrease in the $C_D$ due to the form drag compared to the $C_D$ from the viscous drag. Ultimately, the total drag coefficient also follows a slower rate of decrease than the ${\rm Re}_L^{-0.5}$ trend from the boundary layer theory when compared visually.} In addition, the total viscous drag turns out to be slightly larger than the total viscous drag from the Blasius solution and that can be attributed to the shear stress distribution along the length of the sample being higher, lower, then higher than that of the Blasius solution and thus the differences canceling each other out in the integral.

We can use a similar process to also calculate the lift forces that the sample is experiencing as the result of the small angle of attack and the asymmetry in the flow. For this, the lift is decomposed into the lift due to the pressure difference at the Top and Bottom boundary of the control volume ($L_{\rm pressure}$), then the lift due to pressure and momentum contributions together $L_{\rm CV}$, and ultimately the total lift from the load cell and they are presented in Fig. \ref{fig:Cd_ReL.png}(b). (Note that the lift is normalized by the area of one side of the sample, while drag is normalized by the entire wetted surface area $2Lb$.) Clearly, the slight asymmetry in the flow is able to result in detectable lift values for all the cases with $L_{\rm CV}/D_{\rm CV}  = 0.3$, $0.24$, and $0.49$, and 
$L_{\rm Total}/D_{\rm Total} = 0.26$, $0.19$, and $0.65$ respectively for the 3 cases investigated. However, with the angle of attack being less than $1^{\circ}$, the assumption of $D \approx F_x$ and $L \approx F_y$ is valid as the contributions from $F_y$ to the drag or $F_x$ to the lift would only account for about $0.1\%$ of the total values. 

\section{Conclusion}

\begin{figure*}
    \centering
    \includegraphics[width = 1\textwidth]{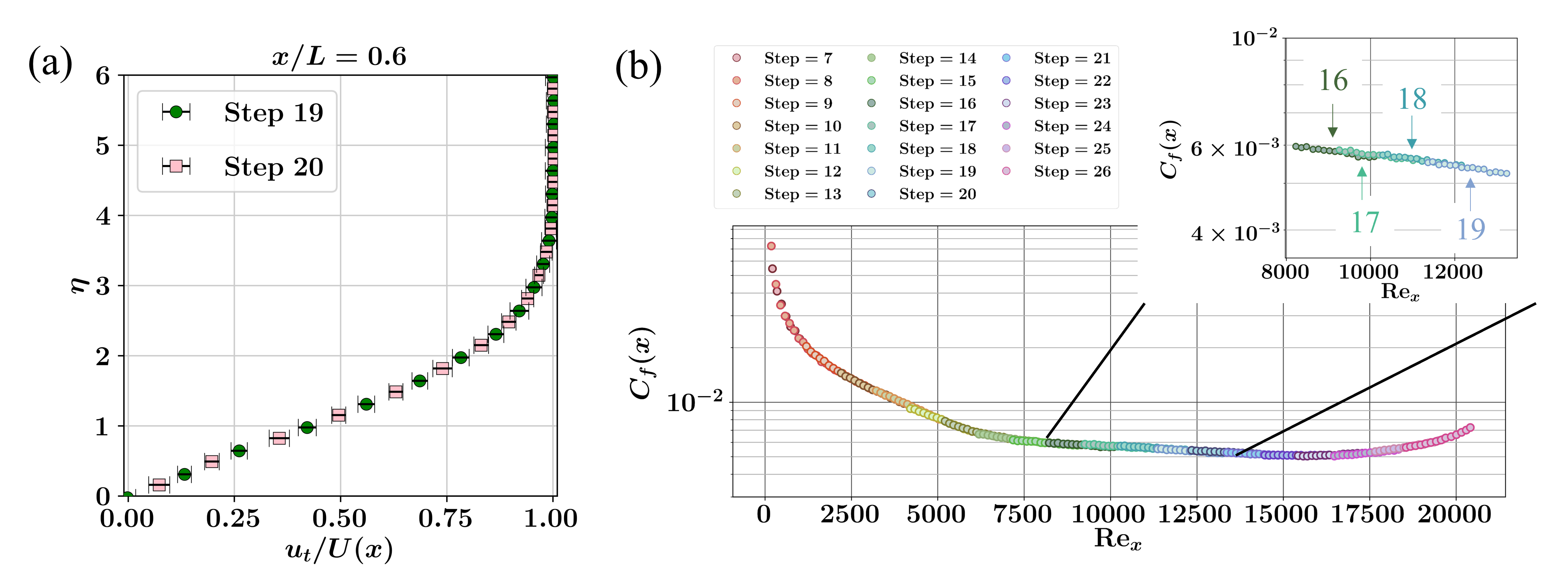}
    \caption{{Demonstration of the similarity of the velocity and shear stress values in the overlapping regions. (a) Velocity profiles extracted from the Front side at $x/L=0.6$ which is in the overlapping region of steps 19 and 20 of the experimental procedure for ${\rm Re}_L = 18,500$. As it can be seen the two velocity profiles are nearly identical. (b) Shear stress distribution on the Front side of the sample operated at ${\rm Re}_L = 18,500$, plotted with cascading colors, each color indicating a different batch of the images from the consecutive-overlapping experiments as shown in the legend. The inset shows a zoomed-in view of the shear stress for folders 16, 17, 18, and 19. As is seen the shear stress values look continuous and the shear stresses in the overlapping regions are nearly the same between the two neighbors.}}
    \label{overlap2}
\end{figure*}

Here we present a cost-effective implementation of the double-light sheet, consecutive-overlapping imaging strategy to perform high-resolution particle image velocimetry experiments with opaque samples that are larger than the field of view of the imaging. We present steps to perform such experiments only using one laser light source and one camera instead of increasing the number of light sources and/or cameras and discuss how a multi-camera simultaneous-overlapping imaging strategy is not feasible with the currently available hardware. Using this method, one can gather data both in the near-wall and the far-field of flow past arbitrary objects and the results can be effectively used for understanding the characteristics of the kinematics and dynamics of different types of external flow problems. 

To demonstrate the capabilities of this technique, we present the results of experiments performed with a slender short flat plate sample that is streamlined at the leading edge at three Reynolds numbers. We present the full view of the velocity distribution and the resulting turbulent kinetic energy and Reynolds shear stress distribution in the wake of the sample. {As it was seen in Sec. \mbox{\ref{Results}}, within the overlapping areas of the imaging steps, we see no difference between results captured from each overlapping step. We also demonstrate that again here in Fig. \mbox{\ref{overlap2}(a)}, showing the velocity profile on the Front side of the case ${\rm Re}_L = 18,500$, at $x/L = 0.6$ which is located in the overlap of images in steps 19 and 20 of the experiments. In addition, we see no gaps or jumps in the shear stress distributions of \mbox{Fig. \ref{fig:Cf_Re.png}} where for instance the overlaps in the shear stress on the Front side of the case ${\rm Re}_L = 18,500$ can be shown using cascading colors as in Fig. \mbox{\ref{overlap2}(b)}.} Access to the high-resolution data in the far field of the samples accompanied by the potential flow model of the flow is used to find the angle of attack of the sample which is less than one degree and not visible to the eye during the experiments.

With access to the high-resolution details of the flow in the boundary layers, we explore the characteristics of the velocity profiles as a function of the local Reynolds number. Employing the family of Falkner-Skan boundary layer solutions and the parameter $m$ in this theory, we fit this model locally to the velocity profiles. Local distribution of this parameter $m$ found as a function of the local Reynolds number shows that for a slim plate with finite thickness and finite length, the behavior of the boundary layer does not follow the Blasius solution. Early at the leading edge, the profiles are more attached to the wall ($m>0$) and then slowly as $m$ is decreased, the profiles move to $m<0$ and they become more detached where slightly after the end of the elliptic leading edge ($x/L \approx 0.34$) the lowest $m$ occurs, and afterward as we move toward the trailing edge, $m$ increases until it moves to $m>0$ where the profiles are then again more attached compared to the Blasius solution continuing until the end of the length of the plate. This behavior is similar on both sides of the plate, however, due to the slight angle of attack the profiles are more attached on the Back of the sample rather than on the Front. 

As a result of this, we can then calculate the local shear stress distribution from the velocity measurements which are very similar to the distribution of $m$. We see that close to the leading edge, the plate experiences shear stress levels more than that captured by the Blasius solution, and then slightly after the end of the elliptical leading edge the shear stress becomes less than the Blasius solution. Then toward the trailing edge, the shear stress takes an increasing trend and goes above the Blasius solution. Again, we see that due to the limited length of the plate, the flow cannot be fully captured by the first-order boundary layer theory. 

In addition, the velocity and shear stress distribution can be used effectively to calculate the total forces exerted on the experimental sample and to decompose the forces into various phenomena at work. The integral of the shear stress offers insight into the total viscous drag force experienced by the sample, while a control volume analysis is used to get a cumulative measure of both viscous and form drag. One can see that a careful assessment of all momentum contributions and pressure forces is required to ensure that the control volume analysis is able to capture the drag force on the sample irrespective of the boundaries chosen. Ultimately, using the load cell measurements, we see that the 3D nature of the sample clearly has some effects on the total forces exerted on the sample compared to what can be captured from the 2D-2C PIV analysis. 

Overall, this experimental platform can be effectively used to study and analyze the near- and far-field flow past objects with complex geometries. Even without idealized flow scenarios and samples, access to the entire flow field allows us to explore various aspects of the flow ranging from extracting a more accurate measure of the angle of the attack of the flow, to better characterization of the boundary layers and the local shear stress distributions and ultimately finding the forces exerted on the sample both using the PIV data and via the load-cell. The ability to collect high-resolution data of such flows will allow us to develop better models as well as more detailed explorations of flows past complex geometries such as but not limited to textured surfaces and roughness elements. {Thus, building upon the previous works \mbox{\cite{michaux2018robopiv, carmer2008evaluation, parikh2023lego, arroyo2008recent, knopp2015investigation, bross2019interaction, sheng2003single}} introducing a few additional optical elements such as beam splitters, and adding a fully computer-controlled position adjustment of the camera to a 2D-2C PIV system, offer a cost-effective way of expanding on the capabilities of a high-resolution 2D-2C PIV technique, and with a similar approach, this procedure can be expanded to PIV experiments at higher Reynolds numbers and turbulent flows as well as those with multi-light sheet strategies of illumination for access to hard-to-reach spaces of more complex geometric samples, and/or 2D camera sweeps for consecutive-overlapping captures of flow past larger objects.}

\section*{Acknowledgement}

This work is supported by the Rowland Fellows program at Harvard University. 
The authors would like to express gratitude to Richard Christopher Stokes for his support with the electronics, undergraduate researchers Lars Caspersen and Mayesha Soshi for their help, Dr. Prasoon Suchandra for discussions regarding the pressure calculations, and Prof. Leah Mendelson for helpful troubleshooting suggestions. 


 

 
 

\bibliography{Ref}


\end{document}